\documentclass[aps,preprint]{revtex4}
\usepackage[version=3]{mhchem}
\usepackage{amsthm}
\usepackage{amssymb,amsmath,amsbsy}
\usepackage[utf8]{inputenc}
\usepackage[Gray,thinqspace,squaren]{SIunits}
\usepackage[normalem]{ulem}
\newcommand{\bs}[1]{{\mathbf #1}}

\renewcommand{\l}{\left}
\renewcommand{\r}{\right}
\newcommand{\mi}{\mathrm i}

\newcommand{\md}{\mathrm d}

\usepackage[usenames,dvipsnames]{color}
\usepackage{graphicx}     % Include figure files
\definecolor{darkblue}{rgb}{0.0,0.0,0.5}

\begin{document}

\title{Instabilities of layers of deposited molecules on chemically
  stripe patterned substrates: {R}idges vs.\ drops}
  
\author{Christoph Honisch}
\affiliation{Institute for Theoretical Physics, University of M\"unster, Wilhelm-Klemm-Str. 9, 48149 M\"unster, Germany}
\author{Te-Sheng Lin}
\affiliation{Department of Applied Mathematics, National Chiao Tung University Hsinchu, 30010 Taiwan}
\author{Andreas Heuer}
\address{Institute for Physical Chemistry, University of M\"unster, Correnstr. 28/30, 48149 M\"unster, Germany}
\affiliation{Center of Nonlinear Science (CeNoS), University of M\"unster, Corrensstr. 2, 48149 M\"unster, Germany}
\affiliation{Center for Multiscale Theory and Computation(CMTC), University of M\"unster, Corrensstr. 40, 48149 M\"unster, Germany}
\author{Uwe Thiele}
\affiliation{Institute for Theoretical Physics, University of M\"unster, Wilhelm-Klemm-Str. 9, 48149 M\"unster, Germany}
\affiliation{Center of Nonlinear Science (CeNoS), University of M\"unster, Corrensstr. 2, 48149 M\"unster, Germany}
\affiliation{Center for Multiscale Theory and Computation(CMTC), University of M\"unster, Corrensstr. 40, 48149 M\"unster, Germany}
%\email{u.thiele@uni-muenster.de}
%\phone{+49\,251\,83\,34939}
\author{Svetlana V. Gurevich}
\affiliation{Institute for Theoretical Physics, University of M\"unster, Wilhelm-Klemm-Str. 9, 48149 M\"unster, Germany}
\affiliation{Center of Nonlinear Science (CeNoS), University of M\"unster, Corrensstr. 2, 48149 M\"unster, Germany}
\affiliation{Center for Multiscale Theory and Computation(CMTC), University of M\"unster, Corrensstr. 40, 48149 M\"unster, Germany}
% \email{gurevics@uni-muenster.de}
\email[Corresponding author: ]{gurevics@uni-muenster.de}  

%  \begin{tocentry}
%  \includegraphics[width=0.5\textwidth]{graph_abstract.eps}
%  \end{tocentry}

\begin{abstract}
  A mesoscopic continuum model is employed to analyse the transport
  mechanisms and structure formation during the redistribution stage
  of deposition experiments where organic molecules are deposited on a
  solid substrate with periodic stripe-like wettability
  patterns. Transversally invariant ridges located on the more
  wettable stripes are identified as very important transient states
  and their linear stability is analysed. It is found that there exist
  two different instability modes that result (i) at large ridge
  volume in the formation of bulges that spill from the more wettable
  stripes onto the less wettable bare substrate and (ii) at small
  ridge volume in the formation of small droplets located on the more
  wettable stripes. These predictions are confirmed by direct
  numerical simulations of the fully nonlinear evolution equation for
  two-dimensional substrates. In addition, the influence of different
  transport mechanisms during redistribution is investigated focusing
  on the cases of convective transport with no-slip at the substrate,
  transport via diffusion in the film bulk and via diffusion at the
  film surface. In particular, it is shown that the transport process
  does neither influence the linear stability thresholds nor the
  sequence of morphologies observed in the time simulation, but only
  the ratio of the time scales of the different process phases.
\end{abstract}

\maketitle

\section{Introduction}
Many coating and surface growth processes through which a solid
substrate is covered by homogeneous or structured layers of the same
or other materials combine deposition and re-distribution stages that
might occur successively or in parallel. Prominent examples are
various homo- and heteroepitaxial surface growth processes
where the re-distribution of material on the substrate occurs through
diffusion of deposited atoms or molecules on the surface of the
substrate \cite{EvTB2006ssr,EiDM2013rmp}.  Often, the material is
first deposited, e.g., via vapor deposition \cite{WZJN2001pms},
molecular beam deposition \cite{Sieg1997pa} or (pulsed) laser
deposition \cite{MiEB2005lc}. Then the adsorbed molecules diffuse
along the surface and form two-dimensional aggregates, complete mono-
or multilayers, and three-dimensional nanoscale structures like
pyramids, holes and mounds - the latter are sometimes called quantum
dots and wells \cite{GoDV2003pre,KoEv2010pd}.

Another example is spin-coating where first a drop of liquid is
deposited on the substrate, before convective motion caused by fast
spinning of the substrate spreads the liquid into a thin film
\cite{WiHD2000jfm,ScRo2004pf}. In the case of solutions or suspensions
the spin coating process is often accompanied by solvent evaporation
and the final coating is a homogeneous or structured layer of the
dried-in solute \cite{ReBD1991jap,ThMP1998prl,MPW2011pf}.  Another
example are dip-coating and related processes where a liquid is
transferred from a bath or other reservoir onto the substrate
\cite{MaOY2003jnr,Thie2014acis}.  These examples all involve a wet
stage where hydrodynamic flows caused by external forces, wettability
and capillarity are important for the redistribution of the material
\cite{WeRu2004arfm}. The flows during the wet stage may be described
employing, e.g., asymptotic models as obtained via long-wave
expansions from the basic equations of hydrodynamics
\cite{oron1997rmp}.

Many of the studied systems involve homogeneous substrates. However, there is
also a growing number of experimental and theoretical studies that
investigate the use of heterogeneous substrates to control the
structure of deposits. Examples include recent deposition experiments
with organic molecules performed by Wang et
al.~\cite{WDWH2011s,WaCh2012acr} on silicon oxide substrates with gold stripes, the
study of dip-coating for chemically micropatterned surfaces
\cite{DTDM2000jap} or dewetting of thin silicon films on nano-patterns
formed by electron beam lithography \cite{BARF2013jap}. For the growth
of quantum dots on heterogeneous substrates see
Ref.~\citenum{LiWY2014pms}.

Static morphologies emerging from capillary and wettability influences
on heterogeneous substrates of various geometries are already studied
in depth, however, studies of the involved dynamics are less frequent:
Morphological transitions of transversally invariant liquid ridges on
two-dimensional striped substrates (i.e., mathematically equivalent to
drops on one-dimensional heterogeneous substrates)  were studied via minimization of
macroscopic interfacial free energies in Ref.~\citenum{LeLi1998prl}
and through bifurcation studies employing mesoscopic free energies
(interface hamiltonians) and time simulations of gradient dynamics on
these energies (equivalent to long-wave hydrodynamic models) in
Refs.~\citenum{BKTB2002pre,thiele2003epje}. In these studies
nonvolatile liquids are considered, i.e., liquid volume and
heterogeneity properties are the main control parameters.
Related steady (one-dimensional) results are obtained in studies of
morphological transitions of wetting films on striped substrates
\cite{BaDP1999el,BaDi2000pre}. There, the main control parameters are
temperature and chemical potential, i.e., the volume of deposited
liquid is a dependent quantity in this grand canonical studies.

Fully two-dimensional situations are also studied:
Ref.~\citenum{thiele2003epje} investigates the linear instability
modes and their time scales for transversally invariant liquid ridges on homogeneous and
heterogeneous (striped) substrates. In particular, they study transversal
instabilities (Plateau-Rayleigh instability) and their coupling to
coarsening instabilities while Ref.~\citenum{BKHT2011pre} also
considers the coupling to depinning modes under lateral driving and
the bifurcation structure for steady two-dimensional states (i.e.,
height profiles depend on both substrate dimensions). The
Plateau-Rayleigh instability is also considered in
Ref.~\citenum{MeRD2008pre} together with the effect of a body force
along the ridge.  Sharma et al.\ present time simulations of dewetting
dynamics for one-dimensional heterogeneous substrates and
two-dimensional substrates with a less wettable square patch
\cite{KaKS2000l}, with stripes \cite{KaSh2001prl} and other
two-dimensional wettability patterns \cite{KaSh2003l}. Droplet
spreading on patterned substrates is also considered \cite{VeSK2011pre}.

Most of the mentioned studies that involve two-dimensional substrates
employ mesoscopic long-wave models (i.e., small gradient expansions)
while Refs.~\citenum{BrLi2002jap,BrKL2005jpcm} determine steady surface profiles
and their stability on the basis of macroscopic interfacial free
energies (note, that this does not allow for a calculation of the time
scales of the instabilities as they result from a balance of
dissipation mechanisms and decrease in free energy). Similar results for
morphological transitions on substrates with micro-grooves are
compared with experiments in Ref.~\citenum{SBKL2005pnasusa}.  Other
experiments concern the pattern-directed dewetting of ultrathin
polymer films that results, e.g., in rows of drops \cite{SFDA2002l}.
Macroscopic experiments with water on chemically heterogeneous
substrates are reported in Ref.~\citenum{GHLL1999s}.  Part of these
works are reviewed in Ref.~\citenum{HeBS2008armr}.

The present study is directly motivated by deposition experiments with
organic molcules performed by Wang et al.~\cite{WDWH2011s,WaCh2012acr}. Several types of
  light-emitting organic molecules (e.g., diferrocene (DiFc); 1,6-
  Bis(2-hydroxyphenol)pyridinel boron
  bis(4-n-butyl-phenyl)-phenyleneamine ((dppy)BTPA);
  N,N'-Di[(N-(3,6-di-tert-butyl-carbazyl))-n-decyl] quinacridone
  (DtCDQA); N,N'-bis-(1-naphyl)-N,N'-diphenyl-1,1'-biphenyl-4,4'-diamine (NPB))
  are deposited through vapour deposition onto SiO$_2$ substrates that
  are prestructured with parallel gold stripes of various width. In
such vapour deposition experiments the ongoing processes often
resemble epitaxial growth of inorganic materials: the molecules
diffuse on the substrate, nucleate at favourable sites and form
growing islands. The prepatterning allows for a control of the
nucleation and growth \cite{WaCh2012acr}.

Although the deposition and redistribution stages often occur
successively, deposition may also still continue when the material is
already being redistributed via diffusion or convective motion. Note
that in particular for highly mobile deposit material, the
  distinction between the two transport processes, diffusion and
  convection, is not sharp when one deals with deposit amounts that
  allow at least locally for multilayer structures. Then one is
  already able to define a velocity profile within the film. In
  consequence, depending on the particular material properties and
  annealing parameters one may expect that depending on the location
  on the substrate transport is dominated by diffusion in a surface
  layer of the deposit (as for solid films discussed, e.g., in
  \cite{SrSa1986japb,MiPS2010rmp}), by diffusion of the bulk layer (as in dynamical density
  functional theory for diffusion of an adsorbate layer \cite{ArRT2010pre}), by
  convection with strong slip at the substrate (as for polymeric
  layers \cite{MWW2005jem}), or by convection without slip (as for the majority
  of liquid layers \cite{oron1997rmp,Thie10}).  In the long-wave models for the time
  evolution for surface profiles of the deposition layers or, more
  generally, the adsorption at the substrate these different transport
  mechanisms are related to power-law mobilities of different powers
  (zero, one, two and three, respectively). A central aim of the
  present work is to clarify in which way a distinction of
  redistribution via the different transport mechanisms is important
  for the observed phenomena.

\begin{figure}[htb]
 \includegraphics[width=.35\textwidth]{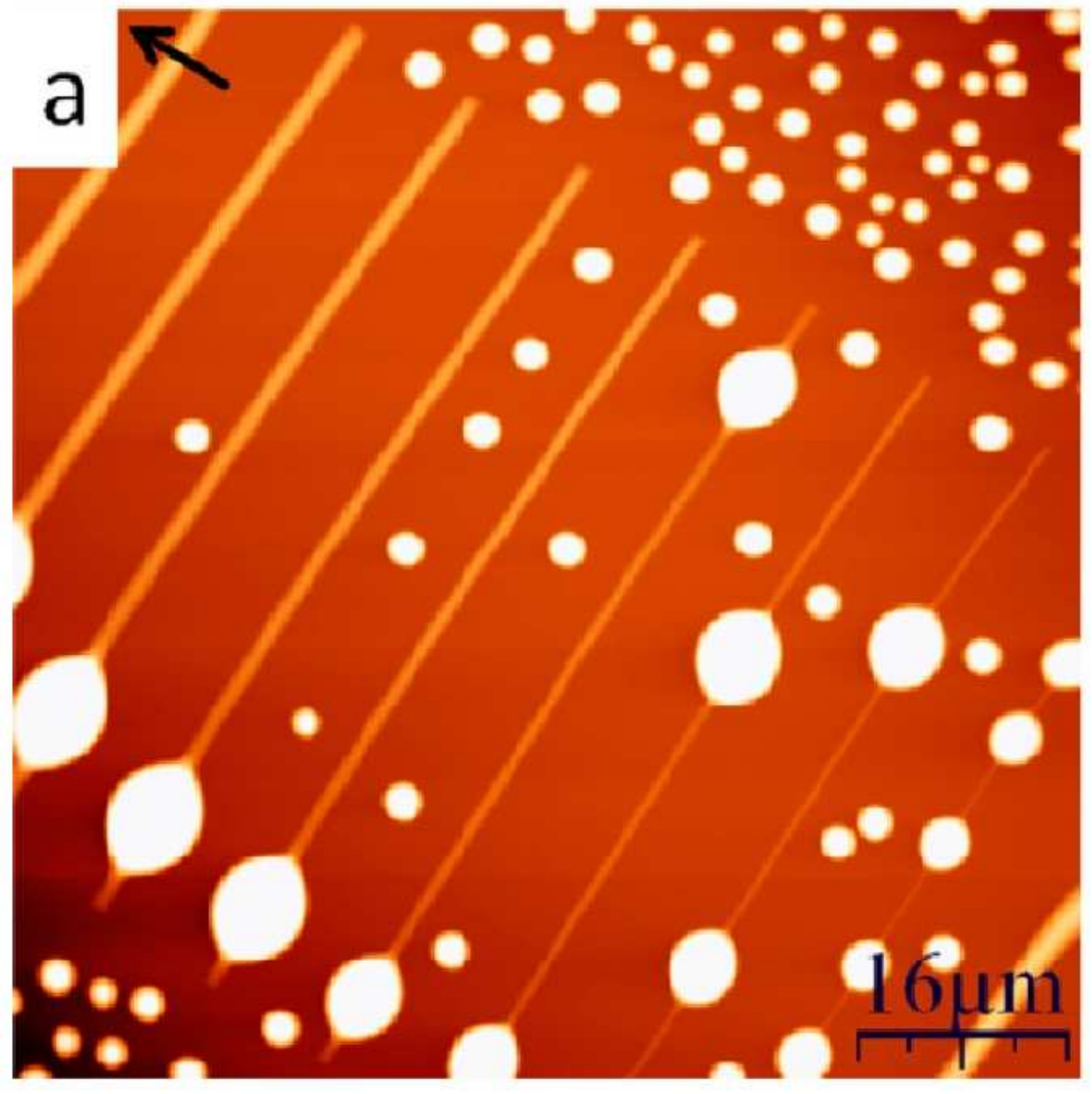}
 \caption{AFM image of DtCDQA molecules deposited on a SiO$_2$
   substrate patterned with Au stripes. The width of the Au stripes is
   increasing from right to left from 0.3$\,\micro$m to
   2.3$\,\micro$m. The mean film thickness is 50\,nm. Since the
   diameter of one molecule is about 1\,nm, this corresponds to
   roughly 50 monolayers. Reprinted with permission from \citenum{WaCh2012acr}. Copyright 2015 American Chemical Society.
   }
 \label{fig:experiments}
\end{figure}

\begin{figure}[htb]
\rotatebox{270}{\includegraphics[width=.4\textwidth]{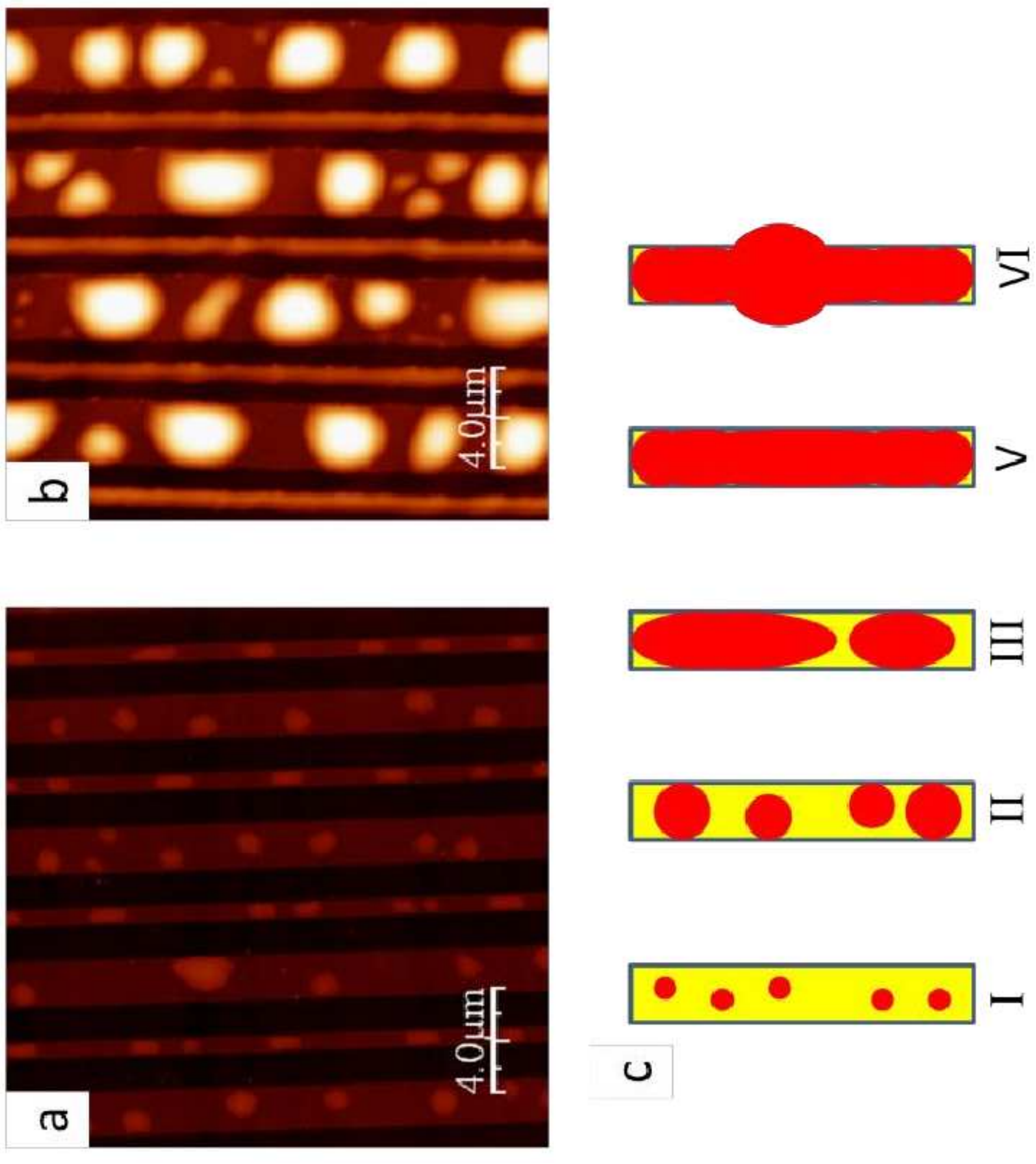}}
 \caption{Different growth regimes during the deposition of DtCDQA molecules on 
SiO$_2$ substrates patterned with Au stripes of alternating width. Panels a) 
and b) show AFM images of 3\,nm and 10\,nm DtCDQA, respectively. Panel c) shows 
sketches of the different growth regimes. For small amounts of deposited molecules, 
small droplets form on the Au lines (I). With increasing amount, larger droplets result 
until droplet edges reach the borders of the stripe (II) which become elongated (III). 
Finally, the whole stripe is covered with a cylindrical ridge (V). At a certain 
amount of molecules, bulges form that also cover part of the SiO$_2$ substrate.
 Picture reproduced from the Supplementary information section of Ref.~\citenum{WDWH2011s}. Copyright (c) 2015 WILEY-VCH Verlag GmbH \& Co. KGaA.}
 \label{fig:experiments2}
\end{figure}

 This is an important question as, e.g., in
  Refs.~\citenum{WDWH2011s,WaCh2012acr} it is discussed that one of
  the employed molecules, DtCDQA, shows a liquid-like behavior while
  the other molecules (as e.g.~NPB) behaves like a solid. In the
  latter, solid case, the height of the rather flat deposits always
  decreases with their width.  In contrast, in the former (liquid )
  case, with increasing amounts of deposited molecules the DtCDQA
  assembles into droplets, elongated drops, molecule stripes,
  cylindrical ridges and ridges with bulges \cite{WDWH2011s}. The
  height of the ridges increases with their width, an effect
  attributed to capillarity (constant contact angles)
  \cite{WaCh2012acr}.  Examples of structures obtained in these
  experiments are reproduced in Figs.~\ref{fig:experiments} and
  \ref{fig:experiments2}.

Note, that the description as 'liquid-like behaviour' in
  Ref.~\citenum{WDLW2013am} is mainly based on the observed shape of
  the deposits (ridges with bulges) but not on an actual observation
  of the transport process or velocity profile across the deposit
  layer. Therefore, one might prefer to describe the behaviour as
  dominated by interfacial tensions, i.e., capillarity. Other
  capillarity-dominated morphologies are obtained for DtCDQA on
  substrates with other patterns, e.g., with a cross-bar structure.
Aspects of these experiments have already been described via
  kinetic Monte Carlo simulations on a lattice \cite{LMWC2012jcp}, in
  particular, the transitions between growth that is localized on the
  gold stripes, the development of bulges that partly cover the bare
  substrate and island formation everywhere on the susbtrate - that
  occur in dependence of the various interaction parameters.

Transformations between liquid ridges with and without bulges are also
observed in macroscopic experiments with water
\cite{GHLL1999s}. There, bulges form when the contact angle for
contact lines pinned at the step in wettability exceeds 90$^{\circ}$.
As the equilibrium contact angles in the experiments with DtCDQA on
Au-striped SiO$_2$ are below 22$^{\circ}$, the macroscopic theory does
not directly apply to the experimental findings.

Here, we employ a mesoscopic continuum description to describe the
deposition experiments. In particular, we are interested in the
stability of transversally invariant ridges of material located on the
more wettable gold stripes and, in general, the role of such ridge
states in the course of the time evolution from a homogeneous
deposited film of molecules towards a final bulge or drop geometry. We
also consider the question of how the dominant redistribution
mechanism (diffusion or convection) influences the evolution pathway.
As our interest is in the static states \textit{and} the dynamic
behaviour and because contact angles are small, we employ a gradient
dynamics in small-slope approximation (i.e., a thin film or long-wave
model) \cite{oron1997rmp,CrMa2009rmp,Thie10}, which describes the
temporal evolution of the film height profile of a thin layer of
material as driven by wettability and capillarity. The heterogeneity
of the substrate is modelled as a chemical heterogeneity that only
affects the wettability similar to
Refs.~\citenum{KaKS2000l,thiele2003epje,KaSh2003l,BKHT2011pre} where a
number of different wetting energies (and therefore Derjaguin
pressures) are used.

The focus of the analysis presented here lies on the different
instabilities of a liquid ridge that result either in the formation of
large bulges that spill from the more wettable stripes  onto the less
wettable bare substrate or in the formation of small droplets on the
more wettable stripes. We perform a linear stability analysis of
steady ridges using continuation techniques
\cite{KrauskopfOsingaEtAl2007,DWCD2014ccp} as outlined in
\citenum{thiele2003epje} and presented in tutorial form in
Ref.\citenum{cenos:lindrop}. In particular, we analyze the influence
of the wettability contrast, the amount of deposited material (mean
film thickness), and the geometry of the stripe pattern on the
stability of liquid ridges. Particular attention is given to an
analysis of the influence of the main transport mechanism in the
redistribution stage.

Our analysis allows us to identify and characterize the different
growth regimes that are experimentally observed in Ref.~\citenum{WDWH2011s}
(here reproduced in Fig.~\ref{fig:experiments2}). Beside the
Rayleigh-Plateau-like instability we describe a second transversal
instability that occurs for very small ridges and show that it is
related to the spinodal dewetting of a thin film on homogeneous
substrates.  The results of the stability analysis are supported by
fully nonlinear time simulations.

The outline of the article is as follows. In Section~\ref{sec:model}
we introduce the gradient dynamics model for the case of a substrate
with a wettability pattern. Next, we explain and illustrate in
Section~\ref{sec:trans} the transversal linear stability analysis
procedure employing a sinusoidal wettability pattern as example. Our
results for a smoothed step-like wettability modulation are given in
Section~\ref{sec:step} while the cases of convective and diffusive
transport are compared in Section~\ref{sec:trans}. In the subsequent
Section~\ref{sec:time} we discuss numerical time simulations for the
two different instabilities again considering various different transport
mechanisms. We conclude and give an outlook in Section~\ref{sec:conc}.

\section{The model}
\label{sec:model}
We employ a mesoscopic continuum model to describe the dynamics of the
redistribution process under the assumption that the slope of the free
surface is everywhere small (long-wave or small gradient approximation
\cite{oron1997rmp,KalliadasisThiele2007}).  In the case of
  purely convective dynamics this assumption allows one to derive an
  asymptotic model from the governing equations of hydrodynamics that
  describes the time evolution of the height profile of a liquid film
  \cite{oron1997rmp,CrMa2009rmp,Thie10}. It reads
\begin{equation} \label{eq:tfe}
  \partial_t\,h(\bs x,t)\,=\,\nabla\cdot\, \left\{Q(h)\,\nabla\,\left[
P(h,\,\bs x)\right] \right\}
\end{equation}
with $\bs x=(x,\,y)^T$ and $\nabla=(\frac{\partial}{\partial_x},\, \frac{\partial}{\partial_y})^T$.
Here, $Q(h)=h^3/(3\eta)$ is the mobility coefficient for a fluid where $\eta$ is the dynamic 
viscosity. The generalized 
pressure $P(h,\,\bs x)$ is given by
\begin{equation} \label{eq:gen_pres}
  P(h,\,\bs x) = -\gamma\Delta h - \varPi(h,\,\bs x)~,
\end{equation}
where $\gamma$ is the liquid-gas surface tension in the Laplace
pressure term and $\varPi(h,\,\bs x)$ is the Derjaguin or 
disjoining pressure~\cite{deGe85,StVe09,Isra11}. For the latter we use \cite{Pism02}
\begin{equation}
  \varPi(h,\, \bs x) = \l( \frac{B}{h^6} - \frac{A}{h^3} \r) (1 + \rho g( \bs x) )~.
\end{equation}
Here the space-dependent term $\rho g(\bs x)$ takes into account the
chemical stripe pattern on the substrate.  The dimensionless parameter
$\rho$ corresponds to the strength of the wettability contrast,
whereas the function $g(\bs x)$ is periodic and describes the geometry
of the stripe pattern.  From now on we use non-dimensional
  variables $\tilde h$, $\tilde x$, $\tilde y$, $\tilde t$ in such a way
  that the non-dimensonal parameters $3\tilde \eta,\,\tilde
  \gamma,\,\tilde A,\, \tilde B$ are all equal to one \cite{note1}. The tilde is
  then dropped in the following.
%\footnote{The film height is scaled by $d = \l(\frac BA\r)^{1/3}$, the $x$ and $y$ coordinates by 
%$L = \sqrt{\frac{\gamma}{A}} d^2$, the effective interface potential by $\kappa = \frac{A}{d^2}$, 
%and the time by $\tau = \frac{3\eta L^2}{\kappa d}$.}
The non-dimensional precursor film height is then uniformly $h_p=1$. Multiplying
the term $1 + \rho g(\bs x)$ to the disjoining pressure keeps $h_p$
constant and only modulates the equilibrium contact angle as
  $\theta_0 =\sqrt{\frac{3}{5}(1 + \rho g(\bs x))}$. Note that
$\theta_0$ is the angle in long-wave scaling, i.e., a small physical
equilibrium contact angle $\theta_e =\epsilon\theta_0$ corresponds to
a long-wave contact angle $\theta_0$ of $O(1)$.

Finally, we note that the hydrodynamic thin film equation corresponds
to a gradient dynamics of the underlying free energy (or interface
Hamiltonian \cite{BEIM2009rmp}). The pressure is given by the
variation $P=\dfrac{\delta F}{\delta h}$ of the free energy $F=\int
\md \bs x \left[\frac{\gamma}{2}|\nabla h|^2 +f(h,\,\bs x)\right]$,
where the first term represents capillarity (interfacial energy of
free surface) and the second term represents position-dependent wettability
(wetting or adhesion energy or binding potential), which is connected
to the disjoining pressure by $\varPi(h,\,\bs x)=-\partial_h f(h,\,\bs
x)$ \cite{Mitlin1993491,Thie10}. Note, that the interpretation
  of the thin film equation~(\ref{eq:tfe}) as a gradient dynamics on
  the free energy functional allows one to go beyond the case of
  convective transport where the mobility is
  $Q(h)=h^3/(3\eta)$ in the case without slip at the substrate
    and $Q(h)=h^2/(3\eta)$ in the case with strong slip at the
    substrate \cite{MWW2005jem}.
  Indeed, by interpreting $h$ as adsorption normalised by a constant
  liquid bulk density, Eq.~(\ref{eq:tfe}) becomes a general kinetic
  equation for the transport of material as driven by interfacial
  energies. In consequence, this allows one to investigate (i)
    transport via diffusion of the entire adsorbed film (then one uses
    the diffusive mobility $Q(h)\propto h$ as in dynamical density
    functional theory \cite{ArRa2004jpag,ArRT2010pre}), and as well
    (ii) transport via diffusion of a surface layer on the deposit as
    in typical solid-on-solid models (then one has a constant mobility
    $Q(h)\propto 1$ \cite{SrSa1986japb,MiPS2010rmp}). Note, that
  similar mobilities have been introduced in a piece-wise thin-film
  model for the formation and motion of droplets on composite
  (meltable) substrates \cite{YoKP2007epje}. A comparative discussion and interpretation of result
    obtained with the various mobilities is provided below.

\section{Linear stability of ridge states}
\subsection{Sinusoidal wettability pattern}
\label{sec:lin}
To illustrate how a continuation method can be employed in the linear
  stability analysis, we first use a simple harmonic
function of one independent variable $x$ as wettability pattern
\begin{equation} \label{eq:g_sin}
  g(\bs x) = \sin\left(\frac{2\pi x}{L_\mathrm{per}}\right),
\end{equation}
i.e., the substrate pattern is invariant in $y$-direction.
The domain size $L$ is chosen equal to the period of the stripe pattern $L_\mathrm{per}$.
First, we determine steady solutions $h_0(x)$ of the one-dimensional system. 
Setting $\partial_t\,h=0$ in Eq.~\eqref{eq:tfe} and integrating twice leads to
\begin{equation} \label{eq:tfe_stat}
  \partial_x^2\,h_0(x) + \varPi(h_0,x) + C = 0\,.
\end{equation}
% \ttuwe{Explain that first integration constant is zero for horizontal substrates.}
The first integration constant corresponds to a net flux into or out
of the integration domain and therefore vanishes in the case of
horizontal substrates without additional lateral driving forces. The second integration constant $C$ is the constant
pressure inside the film that characterizes mechanical equilibrium. In
the following, the constant pressure $C$ is used as a continuation
parameter that plays the role of a Lagrange multiplier which ensures a
constant deposit volume (as the density is assumed to be constant,
here this is equivalent to mass conservation).

Such a one-dimensional solution can be extended homogeneously in
$y$-direction thereby creating a steady solution of the
two-dimensional thin film equation \eqref{eq:tfe} that is
translationally invariant in $y$-direction, i.e., all $y$ derivatives
vanish. To determine the stability of such a solution, we allow for
perturbations that depend in an arbitrary manner on $x$ and are
harmonic in the $y$ direction. We use the ansatz
\begin{equation}
  h(x,y,t) = h_0(x) + \varepsilon\,h_1(x)\,\exp(\beta t + \mi q y)
\end{equation}
where $\beta$ is the growth rate, $q$ is the transversal wavenumber
and $\varepsilon\ll1$.
Introducing this ansatz into equation~\eqref{eq:tfe} leads to lowest order in $\varepsilon$ to the linear eigenvalue equation
\begin{align} \label{eq:eval}
 \beta h_1 = &-Q(h_0) (\partial_x^2-q^2) \nonumber \\ &\times \l[ (\partial_x^2 - q^2)h_1 + 
(\partial_h \varPi(h_0,x))h_1 \r]  \nonumber \\
 &- (\partial_x Q(h_0))\partial_x \l[ (\partial_x^2 - q^2)h_1 + 
(\partial_h \varPi(h_0,x))h_1 \r]  
\end{align}
that has to be solved for unknown $\beta$ and $h_1(x)$.
We solve the time-independent equation~\eqref{eq:tfe_stat} and the
eigenvalue problem \eqref{eq:eval} in parallel using pseudo-arclength
continuation as implemented in the continuation toolbox AUTO-07p
\cite{DoKK1991ijbc,DoedelOldeman2009}. The specific set of equations
treated by AUTO-07p is given in Appendix~\ref{App:AUTO-Eq}.

\begin{figure}[htb]
%\begin{tabular}{l}
 a) 
 %% Plot file: Gruppenplatte/AutoOrdnerChristoph/MyRuns/LinHetDrop/1CH/Plot/pl_BD_r2.plt
\rotatebox{270}{\includegraphics[height=.45\textwidth]{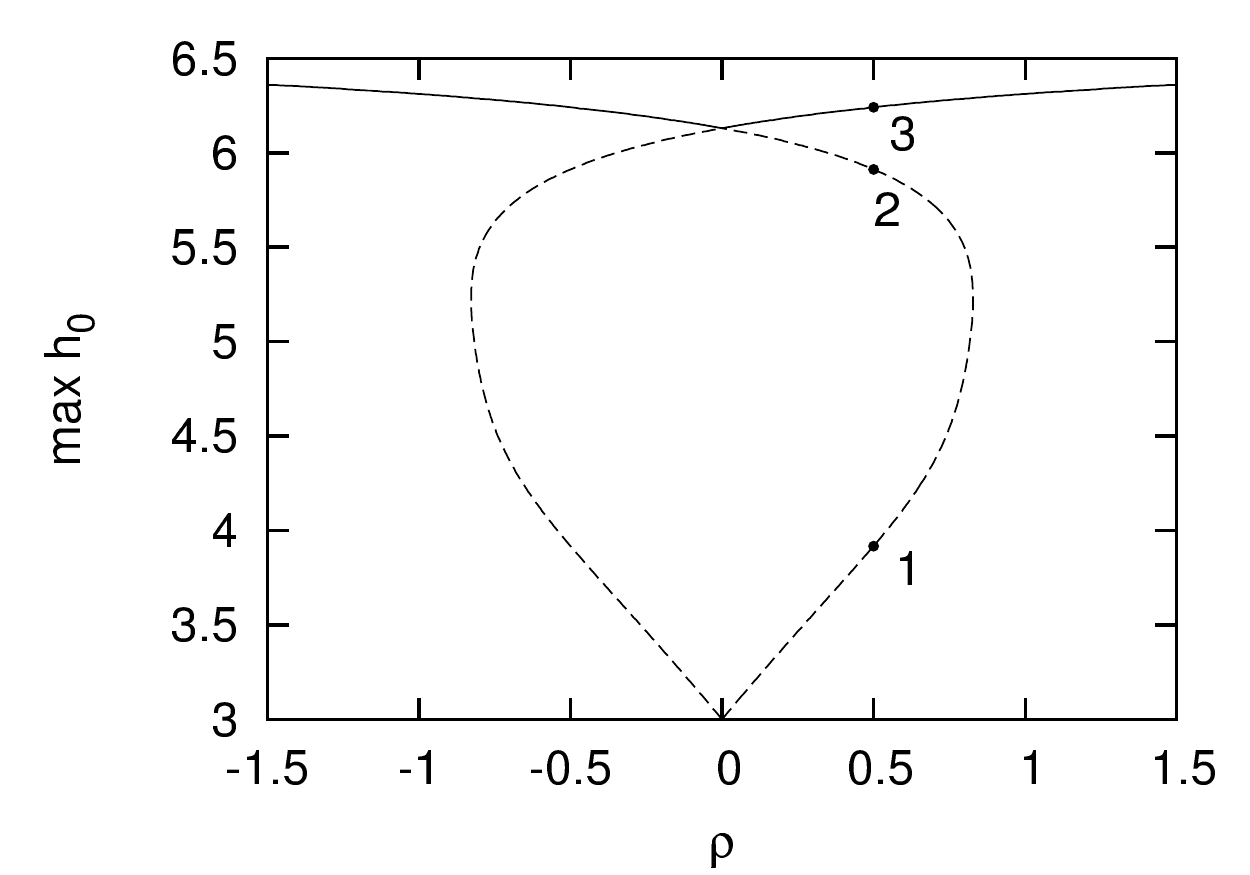}}\\[1ex]
%}\\
 b)
 %% Plot file: Gruppenplatte/AutoOrdnerChristoph/MyRuns/LinHetDrop/1CH/Plot/pl_sol_exmpl.plt
\rotatebox{270}{\includegraphics[height=.45\textwidth]{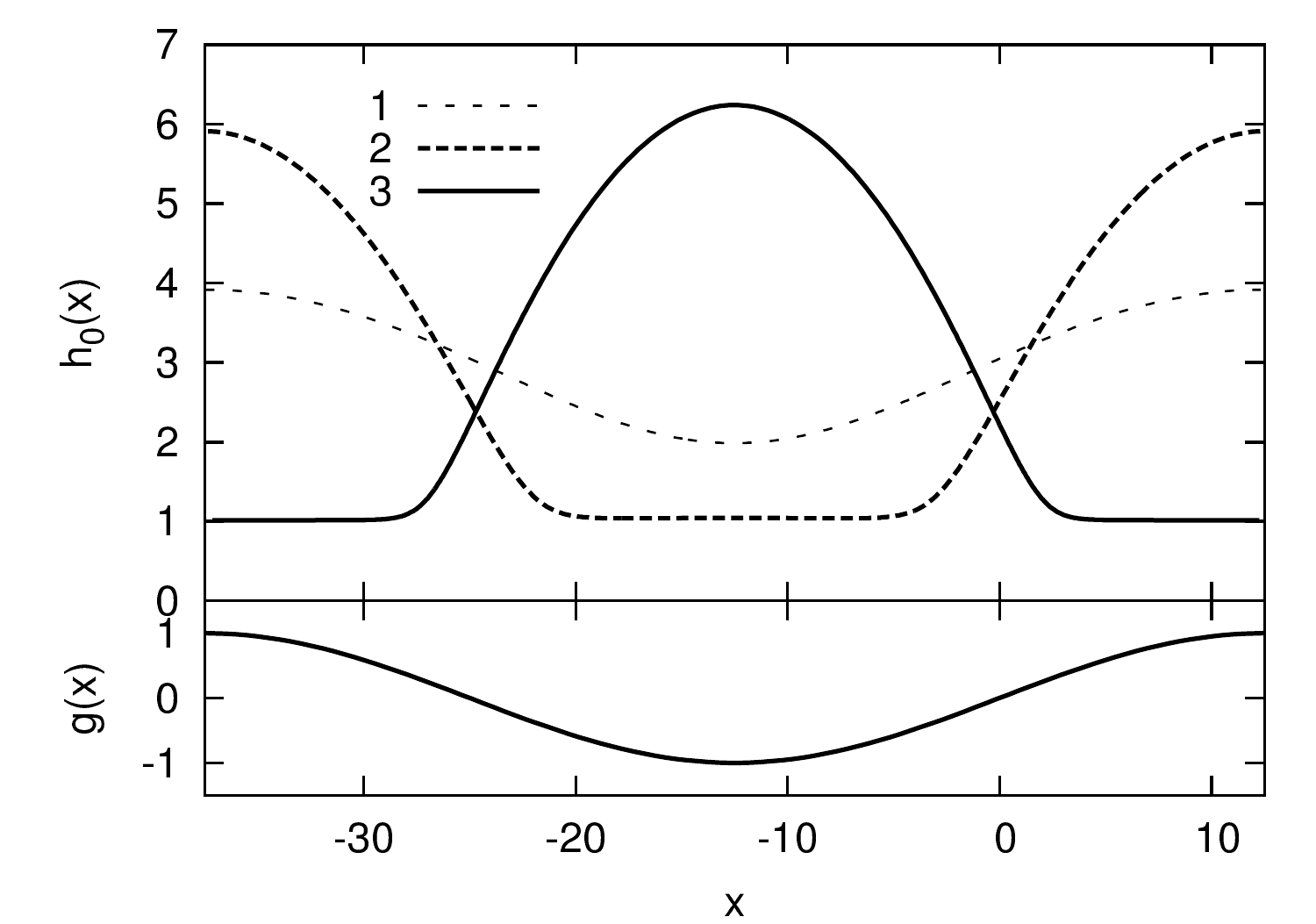}}

%\end{tabular}
  \caption{a) Solution branches obtained from Eqn.~\eqref{eq:tfe_stat}, \eqref{eq:eval} using pseudo-arclength continuation. The solid and dashed parts of the branches represent  stable and unstable solutions, respectively. The parameters are $\bar h=3$ and $L_{per} = 50$. b) Top: Solutions of Eq.~\eqref{eq:tfe_stat} according to the labels in a). Solutions 1 and 2 are unstable. In these cases, more liquid is on the low wettable stripe. Solution 3 represents a stable drop on the more wettable patch. Bottom: The inhomogeneity function $g(x)$ (cf.~Eq.~\eqref{eq:g_sin}). The more (less) wettable area corresponds to $g(x)\approx\,-1$ ($g(x)\approx\,1$).} \label{fig:BD_1}
\end{figure}
 
The desired results are obtained through a number of continuation
runs: We start with the trivial solution $h_0(x) = \bar h = const.$
and $h_1 = 0$, set $\rho,\beta=0$ and $C= -\varPi(\bar h,x)$, whereas
the parameter $q$ is fixed at some small positive value. We continue
this solution as we change the wettability contrast $\rho$. This
yields the solution branches depicted in
Fig.~\ref{fig:BD_1}~a). Figure \ref{fig:BD_1}~b) %\ref{fig:sol_exmpl}
shows three examples of ridge/trench cross-sections according
to the labels in a).

In the next step, we start at a specific point on the stable part of
the solution branch for positive $\rho$ and let the growth rate $\beta$ vary. As $h_1$ is initially zero, and $h_0$ is independent of $\beta$,  and the solution does not change with $\beta$. However, when $\beta$ reaches a value of the discrete
eigenspectrum, a branching point is detected as a solution branch with
$h_1 \neq 0$ bifurcates vertically. In the next run one follows this
new branch, effectively only changing the amplitude of $h_1$ at fixed
$\beta$. The norm $||h_1||$ is used as an additional continuation
parameter and the run is stopped when $||h_1||=1$.

\begin{figure}[htb]
  a)
  %% Plot file: ~/Dropbox/AUTO/MyRuns/LinHetDrop/1CH/Plot/pl_emode.plt
  \rotatebox{270}{\includegraphics[height=.45\textwidth]{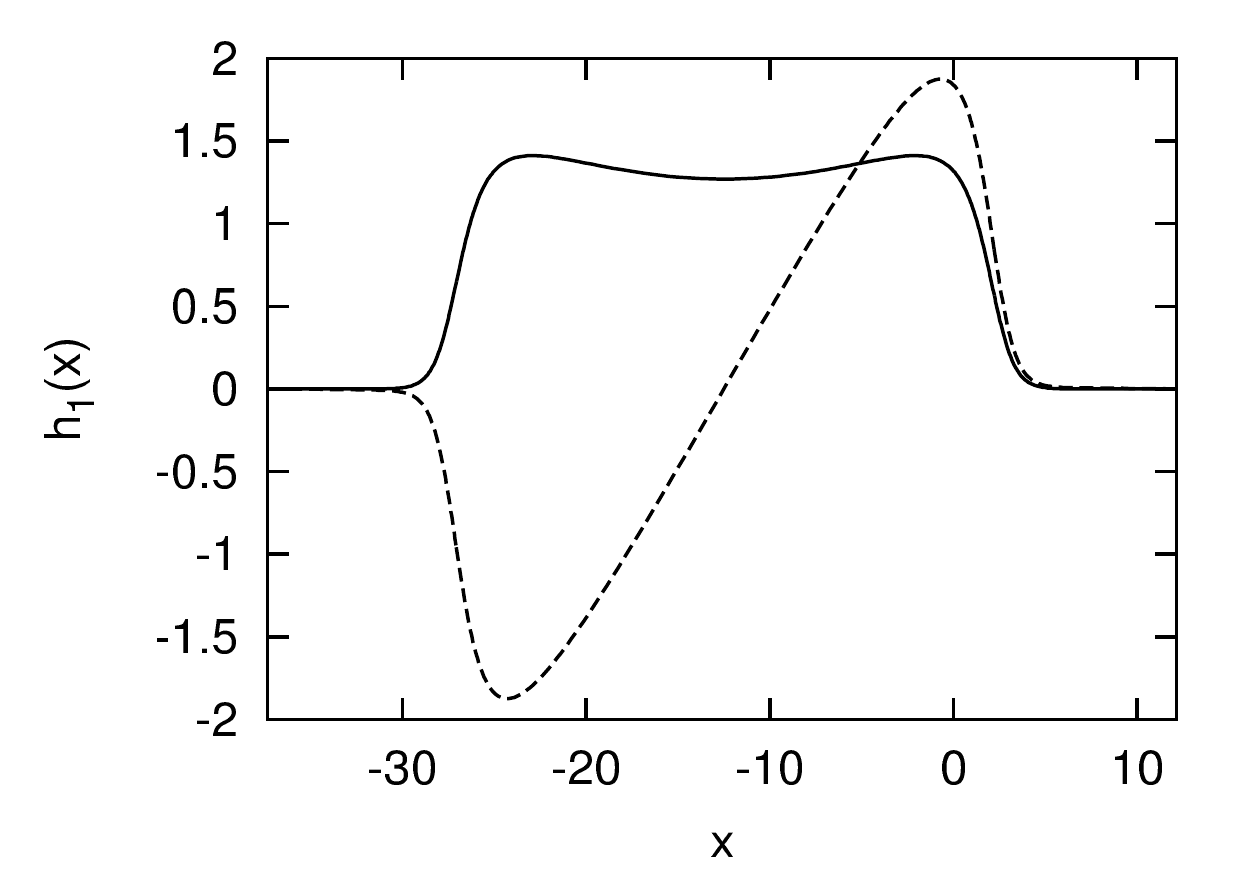}}\\
  b)
  %% Plot file: ~/Dropbox/AUTO/MyRuns/LinHetDrop/1CH/Plot/pl_disp_rel.plt
  \rotatebox{270}{\includegraphics[height=.45\textwidth]{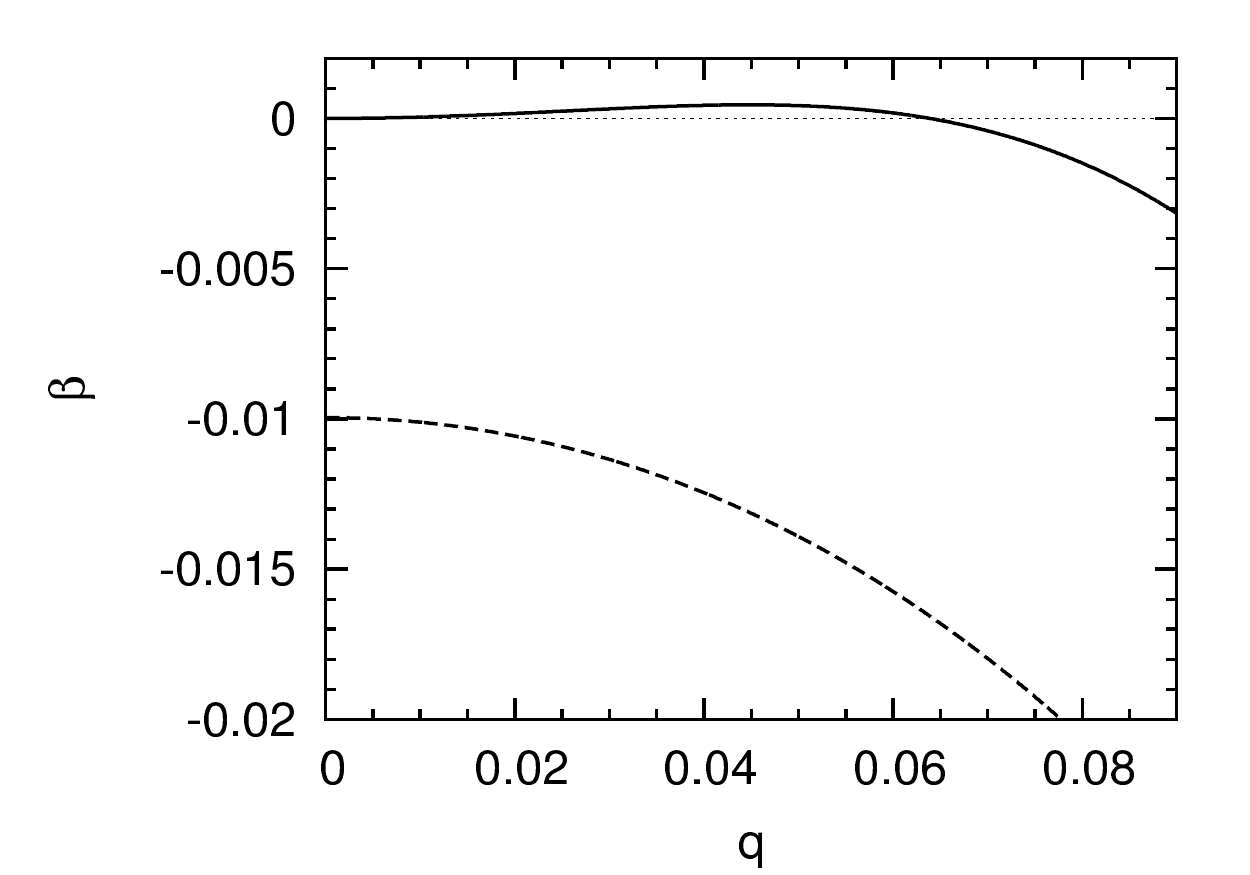}}
\caption{a) The two most important eigenmodes of Eq.~\eqref{eq:eval}. The solid curve is an unstable varicose
mode, the dashed curve is a stable zigzag mode. ($\rho=0.5$, $\bar h=3$). b) Dispersion relations of the two critical modes shown in a). The varicose mode (solid line) has a band of finite wavelengths with positive growth rates. Therefore the analyzed ridge is linearly unstable w.r.t. this mode, whereas the zigzag mode (dashed) is always stable for the system in question.} \label{fig:emode}
\end{figure}
%
% \begin{figure}[htb]
% %% Plot file: ~/Dropbox/AUTO/MyRuns/LinHetDrop/1CH/Plot/pl_emode.plt
%   \rotatebox{270}{\includegraphics[height=.45\textwidth]{emode}}
%  \caption{The two most important eigenmodes \ttsveta{of Eq.~\eqref{eq:eval}. The solid curve is an unstable varicose
% mode, the dashed curve is a stable zigzag mode.} The corresponding dispersion
%  relations are plotted in Fig.~\ref{fig:DR}. ($\rho=0.5$, $\bar h=3$).} \label{fig:emode}
% \end{figure}
% \begin{figure}[htb]
% %% Plot file: ~/Dropbox/AUTO/MyRuns/LinHetDrop/1CH/Plot/pl_disp_rel.plt
%   \rotatebox{270}{\includegraphics[height=.45\textwidth]{disp_rel}}
%   \caption{Dispersion relations of the two \ttsveta{critical} modes shown in Fig.~\ref{fig:emode} \sout{with 
% corresponding line styles.} The varicose mode (solid line) has a band of finite wavelengths 
% with positive growth rates. Therefore the analyzed ridge is linearly unstable \ttsveta{w.r.t. this mode, whereas the zigzag mode (dashed) is always stable for the system in question.}} \label{fig:DR}
% \end{figure}

Figure~\ref{fig:emode}~a) shows the first two eigenmodes of
Eq.~\eqref{eq:eval}. The symmetric eigenmode (solid line) corresponds
to a so-called varicose mode and is unstable, whereas the
antisymmetric zigzag eigenmode (dashed line) is stable
\cite{BrRe1992l, thiele2003epje}.  In the unstable $q$-range the
ridge is linearly unstable w.r.t.\ a Rayleigh-Plateau instability
(cf.~Refs.~\citenum{thiele2003epje,BKHT2011pre} for related thin film
systems).

% \begin{figure}[htb]
% %% Plot file: ~/Dropbox/AUTO/MyRuns/LinHetDrop/1CH/Plot/pl_emode.plt
%   \rotatebox{270}{\includegraphics[height=.45\textwidth]{emode}}
%   \caption{The two most important eigenmodes. The corresponding dispersion
%     relations are plotted in Fig.~\ref{fig:DR}. 
% ($\rho=0.5$, $\bar h=3$).} \label{fig:emode}
% \end{figure}

Now one fixes the norm $||h_1||=1$ and continues the eigenvalue
problem in the parameters $\beta$ and $q$ to directly compute the
dispersion relation $\beta(q)$ for each eigenmode. The results for
both, varicose and zigzag mode, are shown in
Fig.~\ref{fig:emode}~b). The varicose mode has a finite band of
unstable wavenumbers $0<q<q_c$ similar to the case of a homogeneous
substrate.  The growth rate at $q=0$ is always zero for the varicose
mode: $\beta(q=0)=0$ corresponding to mass conservation.
The zigzag mode is always stable for the present system. On a homogenous
substrate $\beta(q=0)=0$ reflects the translational invariance in
$x$ direction\cite{thiele2003epje}. In contrast, here this invariance is
broken by the stripe wettability pattern implying that $\beta(q=0)\neq0$.
Note, that the zigzag mode might become unstable under driving in
$x$ direction, e.g., on an incline \cite{ThKn03}.

\begin{figure}[htb]
%% Plot file: ~/Dropbox/AUTO/MyRuns/LinHetDrop/1CH/Plot/pl_phase_portrait.plt
  \rotatebox{270}{\includegraphics[height=.45\textwidth]{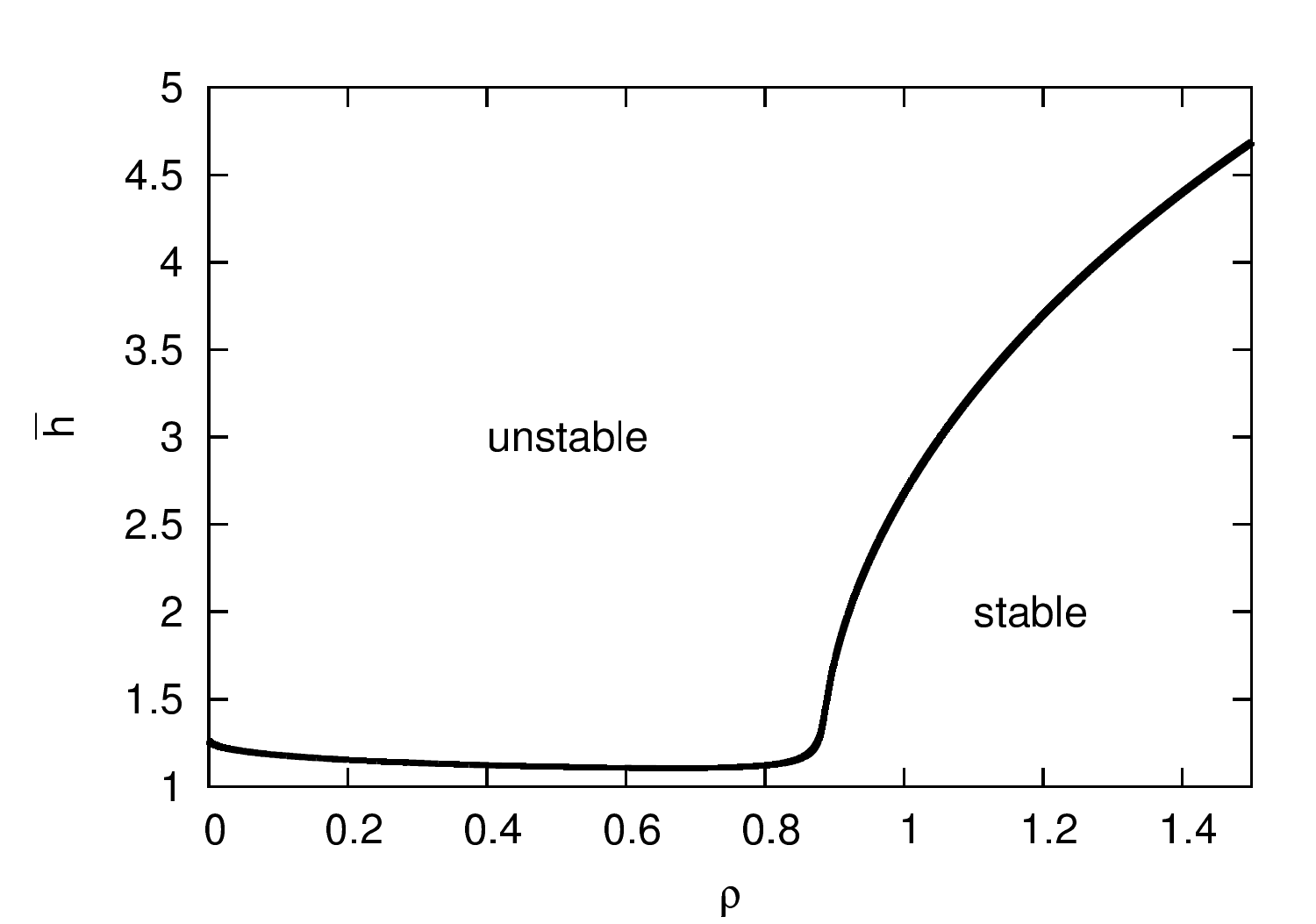}}
  \caption{Linear stability diagram in the $[\rho, \bar h]$ plane. The solid line shows 
the linear stability threshold for a homogeneous ridge solution. For parameters below and 
above the curve, ridges are linearly stable and unstable, respectively.} \label{fig:phase_portrait}
\end{figure}

We employ AUTO-07p in such a way that the maximum of the dispersion
relation corresponds to a saddle-node bifurcation of the curve
$q(\beta)$. With this trick one is able to follow the maximum of the
dispersion relation when varying some other parameter through a fold
continuation. We follow the maximum of $\beta(q)$ for the varicose
mode varying, e.g., $\bar h$, until $\beta \approx 0$. A final
continuation follows this point in the plane spanned by $\bar h$ and
$\rho$ and directly gives the instability threshold. This allows us to
obtain the stability diagram for ridges presented in
Fig.~\ref{fig:phase_portrait}.

For large wettability contrasts $\rho$, the stability threshold with respect to $\bar h$ 
increases monotonically with $\rho$. This is the behavior we expect from the experimental 
findings. For small $\rho$, however, this is no longer true. We will discuss this surprising 
result at the end of the next section.

\subsection{Smoothed-step stripe-like inhomogeneities}
\label{sec:step}

In the previous section we have introduced our methodology using a simple sinusoidal wettability
modulation. Now we use a spatial modulation of the disjoining pressure that realistically reflects the experimentally employed chemical stripe
pattern. We use smoothed steps employing
\begin{align}\label{eq:spg}
%  g(x) = &\tanh \l[ \frac{1}{l_s}\l( -\text{frac} \l( \frac{x}{L_{per}} \r) +x_A  \r)  \r] \nonumber \\
%  &\times \tanh \l[ \frac{1}{l_s}\l( -\text{frac} \l( \frac{x}{L_{per}} \r)+(1 -x_A)  \r)  \r]~,
  g(x) = &\tanh \l[ \frac{1}{l_s}\l( x_A-\l( \frac{x}{L_{per}} \r)   \r)  \r] \nonumber \\
  &\times \tanh \l[ \frac{1}{l_s}\l((1 -x_A) -\l( \frac{x}{L_{per}} \r)  \r)  \r]~,
\end{align}
%where frac($y) = y -$floor($y$) is the fractional function. 
Figure~\ref{fig:sol2}~c) shows an example of the function
$g(x)$. The parameter $L_{per}$ is the spatial period as before, $l_s
L_{per}$ is the width of the transition region between the more
wettable stripe (MWS) and the less wettable stripe (LWS).  That is,
$l_s$ defines the sharpness of the wettability contrast. The MWS
($g(x)\approx -1$) starts at $x_A L_{per}$ and ends at
$(1-x_A)L_{per}$ in each period. The smaller $x_A$ is, the wider is
the MWS.
%
% \begin{figure}[htb]
% %% Plot file: ./pl_g.plt
%   \rotatebox{270}{\includegraphics[height=.45\textwidth]{fun_g}}
%   \caption{\ttsveta{The spatial modulation of the disjoining pressure} \sout{The function} $g(x)$ \ttsveta{\eqref{eq:spg}} for parameters $L_{per} = 50$, $l_s=0.03$ and $x_A=0.3$.} \label{fig:fun_g}
% \end{figure}

\begin{figure}[htb]
%% Plot file: ~/Dropbox/AUTO/MyRuns/LinHetDrop/1CH/TANH/Plot/pl_bif_r1r2.plt
  \rotatebox{270}{\includegraphics[height=.45\textwidth]{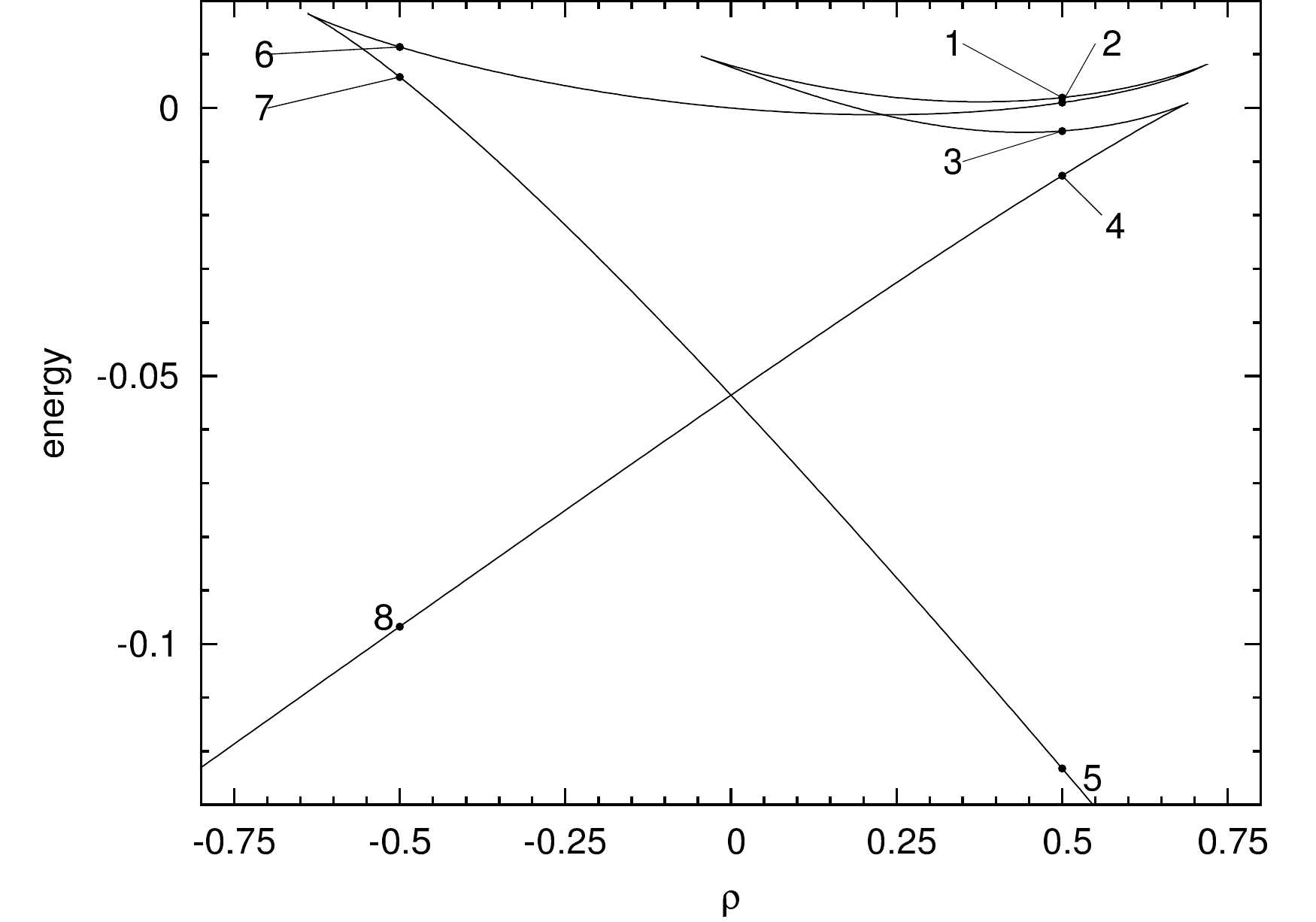}}
  \caption{Solution branches obtained by a first continuation in the wettability contrast $\rho$.
In contrast to Fig.~\ref{fig:BD_1}, we use the energy as a solution measure in order to assess
the stability of the solutions. The solutions corresponding to the labels $1-8$
are shown in Fig.~\ref{fig:sol2}. The parameters are $L_{per} = 50$, $l_s=0.03$,
$\bar h=3$ and $x_A=0.3$.
} \label{fig:BD_2}
\end{figure}

\begin{figure}[htb]
 \includegraphics[height=.55\textwidth]{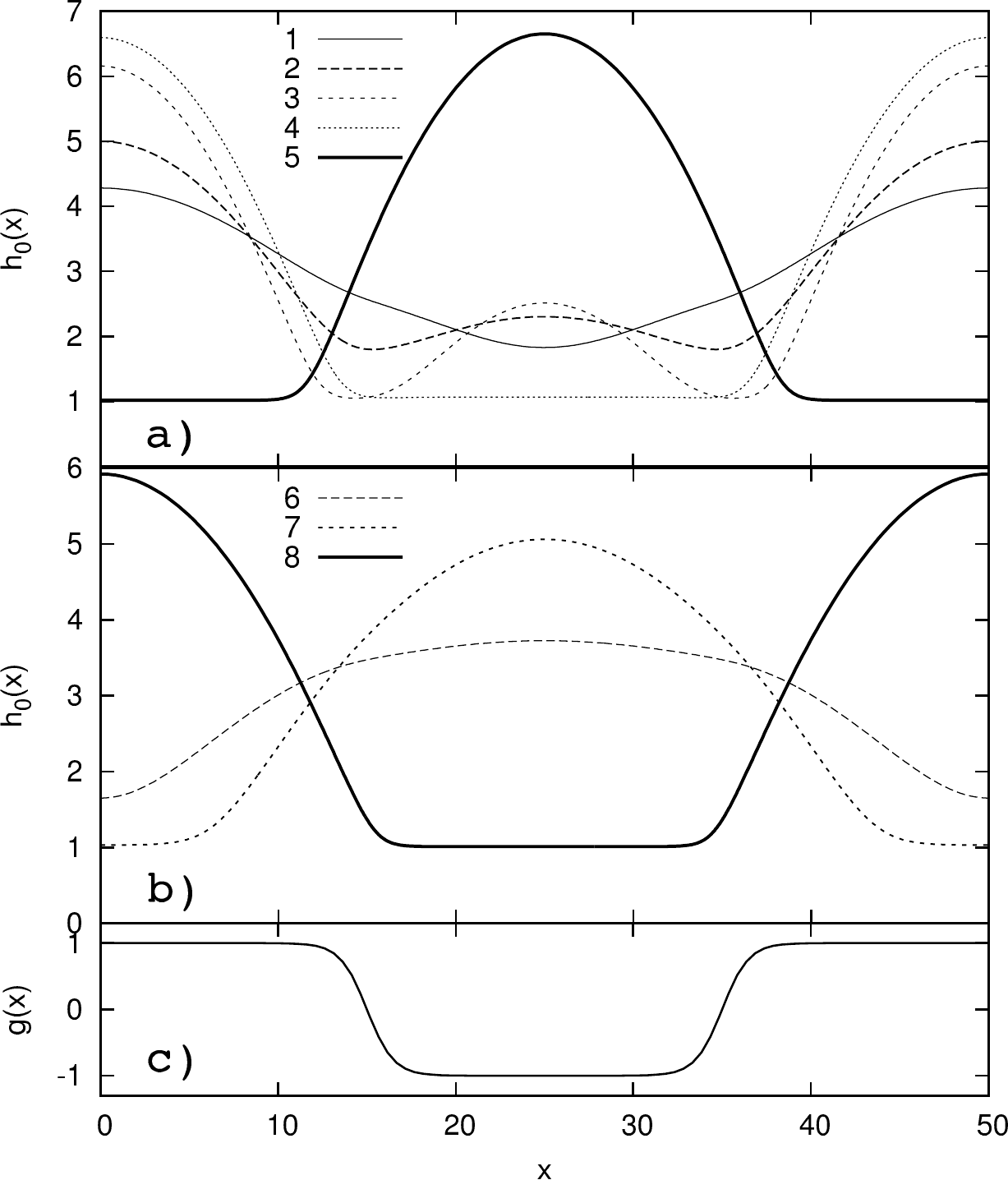}
\caption{Steady solutions corresponding to the labels a) $1-5$  and b) $6-8$ of Fig.~\ref{fig:BD_2} for 
$\rho=0.5$ ($\rho=-0.5$), $L_{per} = 50$, $l_s=0.03$, $\bar h=3$ and $x_A=0.3$. As $\rho$ 
is positive (negative), the more wettable area corresponds to the
region where $g(x)\approx -1$ ($g(x)\approx 1$). Here, solutions five
(a)) and eight (b)) are the only stable solutions. c) The
inhomogeneity function $g(x)$, given by Eq.~\eqref{eq:spg}.}
\label{fig:sol2}
\end{figure}

We start with a domain size equal to the period of the stripe pattern $L_{per}=50$ and set $x_A=0.3$ so that the MWS is thinner than the LWS. The parameter $l_s$ is set 
to $0.03$. As in the previous section, the starting solution is the trivial one $h_0=\bar h=3$. 
We continue this solution as $\rho$ is varied and obtain the solution branch shown in 
Fig.~\ref{fig:BD_2}. Since the MWS is thinner than the LWS, the symmetry between positive 
and negative $\rho$, which could be seen in Fig.~\ref{fig:BD_1}~a) for
sinusoidal wettability patterns, is broken. Figure
\ref{fig:sol2} shows the solutions for $\rho=0.5$ (Fig.~\ref{fig:sol2}~a)) and $\rho=-0.5$ (Fig.~\ref{fig:sol2}~b)) , 
respectively. For positive $\rho$ we find a larger variety of unstable solutions, e.g., 
configurations with a small drop on the MWS and a larger drop on the LWS.


\begin{figure}[htb]
%% Plot file: ~/Dropbox/AUTO/MyRuns/LinHetDrop/1CH/TANH/Plot/pl_phase_portrait2.plt
  \rotatebox{270}{\includegraphics[height=.7\textwidth]{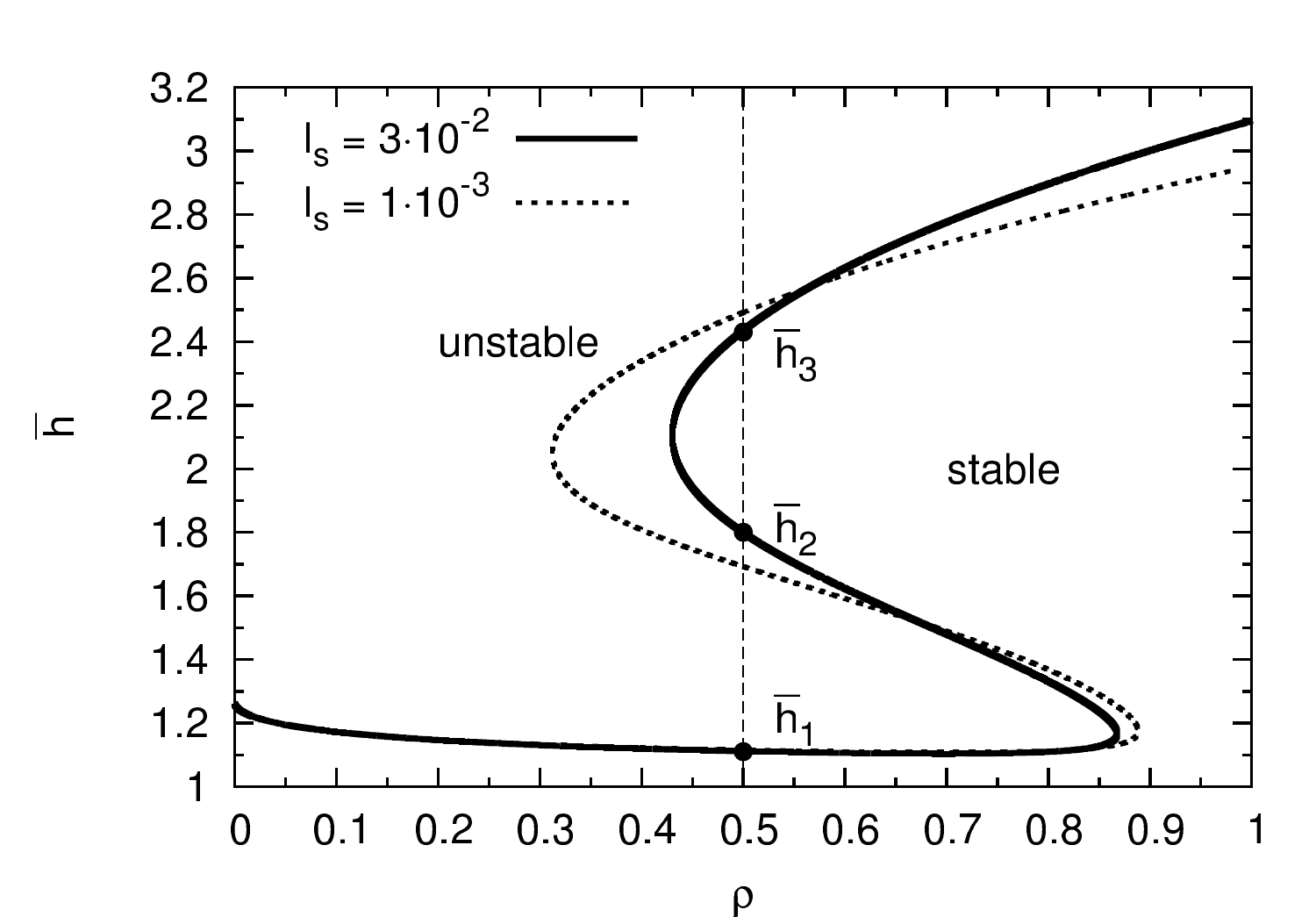}}
  \caption{Linear stability diagram in the $[\rho, \bar h]$ plane for
    steep wettability gradients between the stripes. In particular, $l_s = 0.03$ 
(solid line) and $l_s=0.001$ (dotted line).  For $\rho = 0.5$ there are three stability thresholds $\bar h_1,\,\bar h_2$ and $\bar h_3$.
} \label{fig:phase_portrait2}
\end{figure}

As in the previous section, we  directly compute the linear stability diagram on
the $[\rho, \bar h]$ plane. The result is shown as the solid line in
Fig.~\ref{fig:phase_portrait2}.
In contrast to the example with the sinusoidal wettability modulation
(Fig.~\ref{fig:phase_portrait}), the curve has an essentially
non-monotonous form.  In the region $0.43\lesssim \rho \lesssim 0.87$,
there are four different stability regions that deserve further
investigation. To this end, we consider the system for a constant
$\rho=0.5$.

%\subsection{Detailed analysis for $\rho=0.5$}

\begin{figure}[htb]
 \begin{tabular}{l}
  a)\\
  %% Plot file: ~/Dropbox/AUTO/MyRuns/LinHetDrop/1CH/TANH/Plot/pl_sol_trans.plt
  \rotatebox{270}{\includegraphics[height=.35\textwidth]{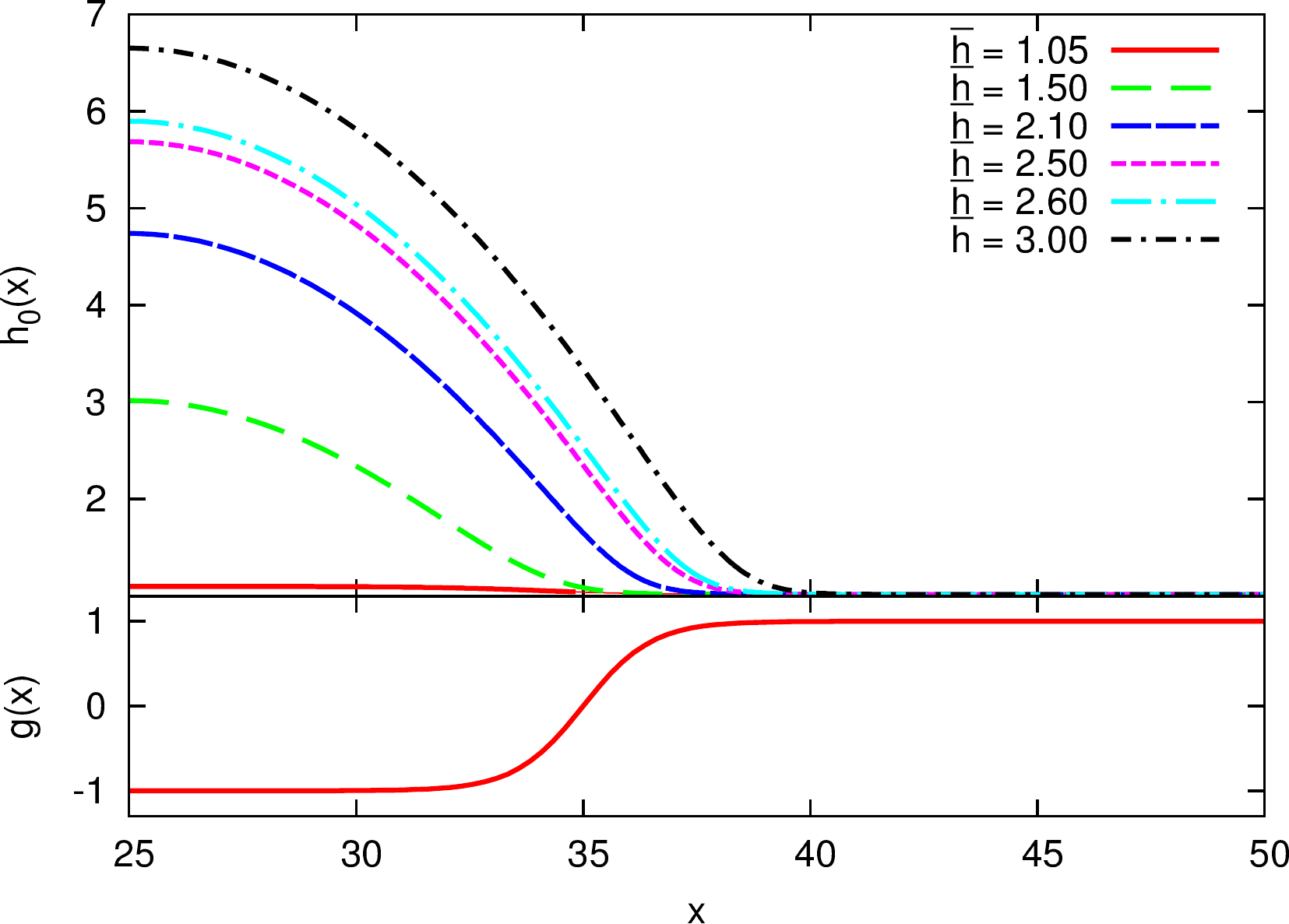}}\\
  b)\\
  %% Plot file: ~/Dropbox/AUTO/MyRuns/LinHetDrop/1CH/TANH/Plot/pl_ef_trans.plt
  \rotatebox{270}{\includegraphics[height=.35\textwidth]{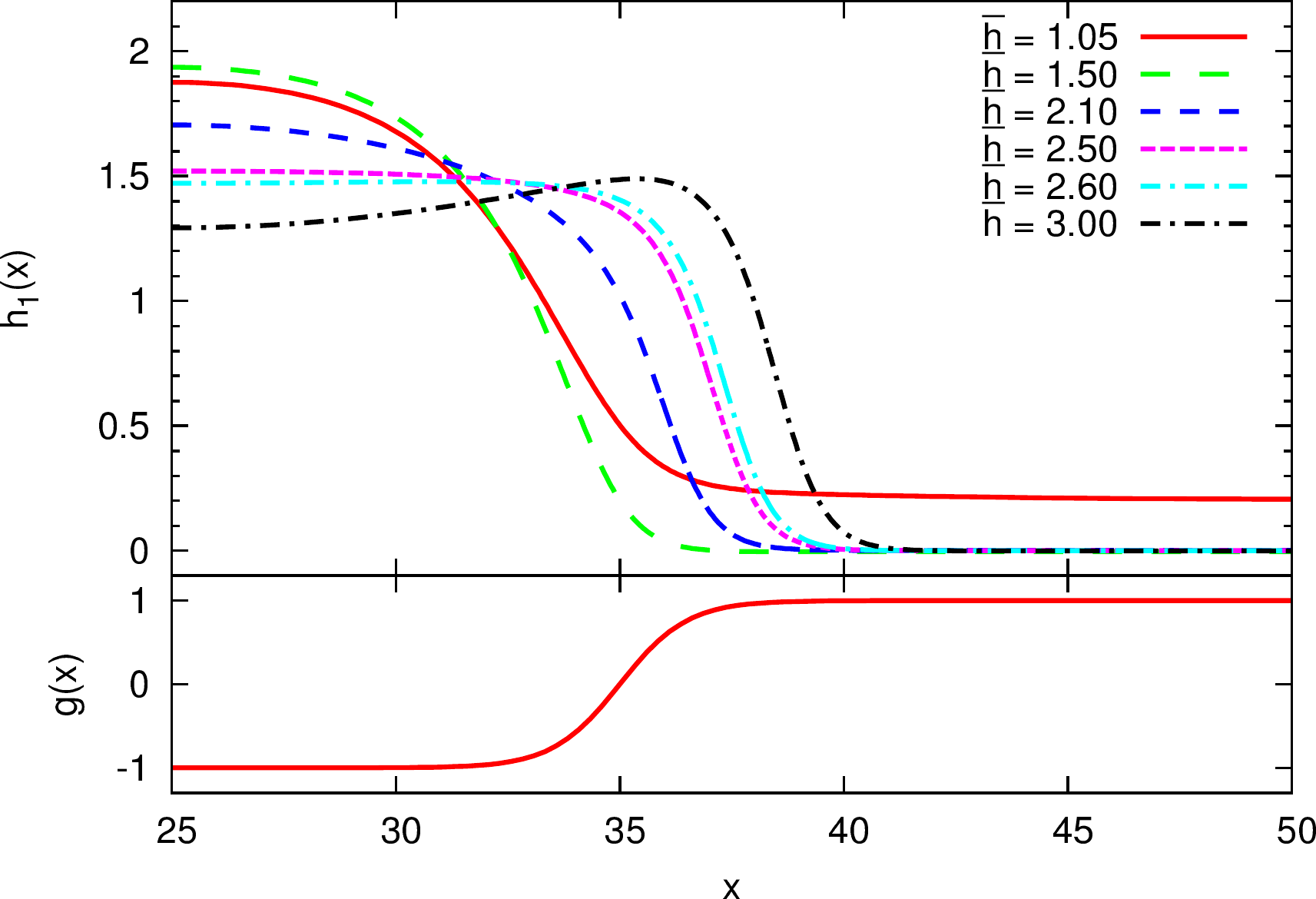}}
 \end{tabular}
\caption{Top panels: Stationary solutions $h_0(x)$ (a)) and critical eigenfunction $h_1(x)$ (b)) for $\rho=0.5$ and different mean film heights $\bar h$. At $2.5<\bar h<2.6$, the transition between a unimodal and a bimodal shape occurs. Bottom panels: The inhomogeneity function $g(x)$, given by Eq.~\eqref{eq:spg}.}\label{fig:sol_trans}
\end{figure}

\begin{figure}[htb]
\includegraphics[height=.29\textwidth]{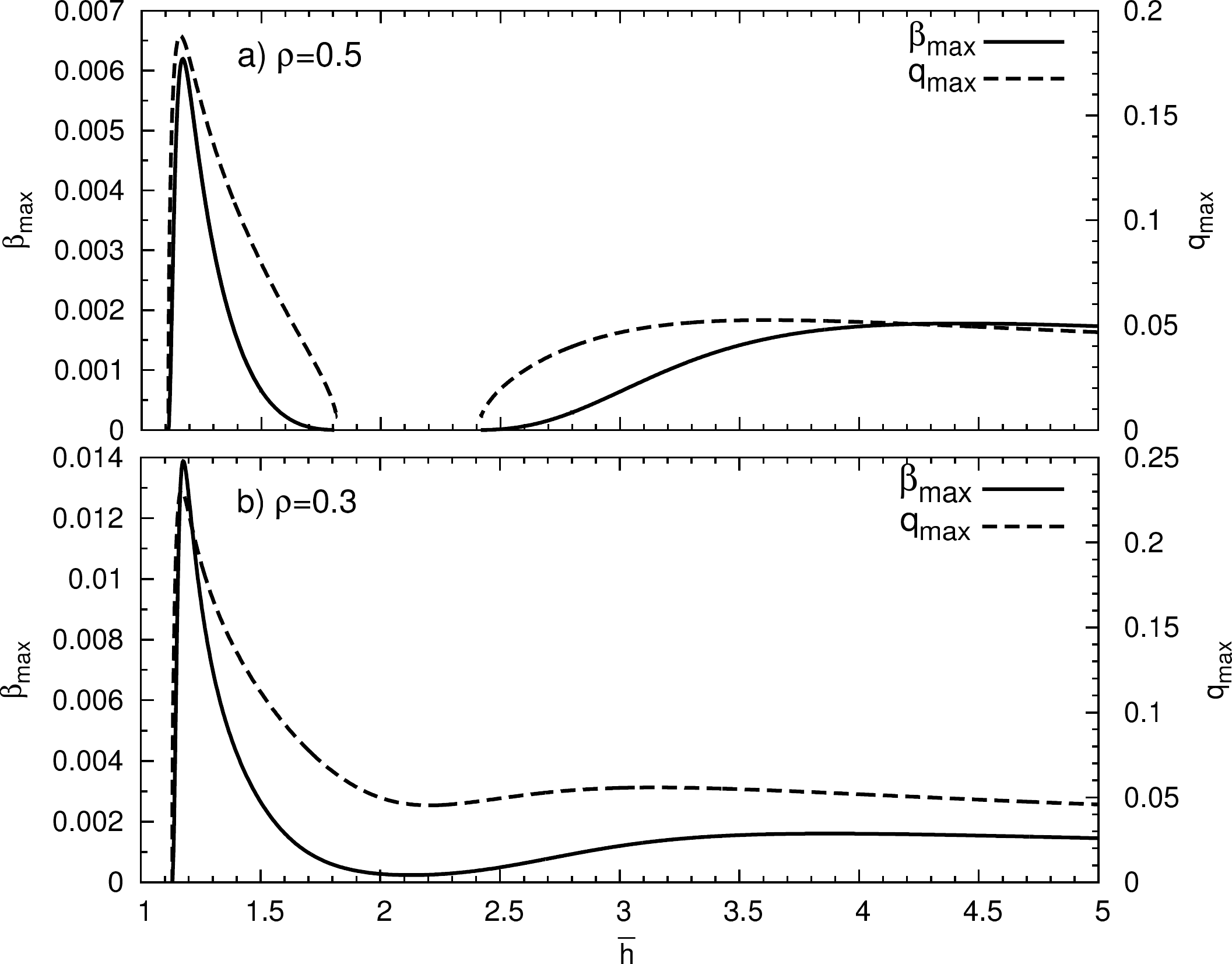}
 \caption{Maximal growth rates $\beta_{\text{max}}$ and the corresponding $q_{\text{max}}$, 
which are defined in Eq.~\eqref{eq:def_betamax}, against $\bar h$ for a) $\rho=0.5$ and b) 
$\rho=0.3$.} \label{fig:beta_hmean}
\end{figure}

For $\rho=0.5$ we observe three critical values of the mean 
film thickness $\bar h$, $\bar h_1 \approx 1.1$, 
$\bar h_2 \approx 1.8$ and $\bar h_3 \approx 2.4$. We continue 
the stable one-dimensional solution $h_0(x)$ and its critical eigenfunction $h_1(x)$ in the mean 
film thickness $\bar h$. Figure \ref{fig:sol_trans} shows the 
obtained profiles $h_0(x)$ and eigenmodes $h_1(x)$, respectively. In Fig.~\ref{fig:beta_hmean} a), 
we give the maximal growth rates
\begin{equation} \label{eq:def_betamax}
  \beta_{\text{max}}=\max_q \beta(q)=\beta(q_{\text{max}})
\end{equation}
and the corresponding $q_{\text{max}}$ in the unstable regions $\bar h_1 < \bar h < \bar h_2$ 
and $\bar h > \bar h_3$. In the stable regions $\bar h < \bar h_1$ and $\bar h_2 < \bar h < \bar h_3$, 
$\beta_{\text{max}}=q_{\text{max}}=0$.

For very thin films ($\bar h< \bar h_1$), the one-dimensional
 solutions are not yet droplets.  They are better
described by piecewise flat films with a higher mean film thickness on
the MWS (cf.~solution for $\bar h=1.05$ in
Fig.~\ref{fig:sol_trans}~a)). The flat film pieces are sufficiently thin to be  below the instability threshold for 
spinodal dewetting on respective  homogeneous
substrates. Therefore, the piecewise flat film solutions are stable,
also in two dimensions.

\begin{figure}
  \rotatebox{270}{\includegraphics[height=.49\textwidth]{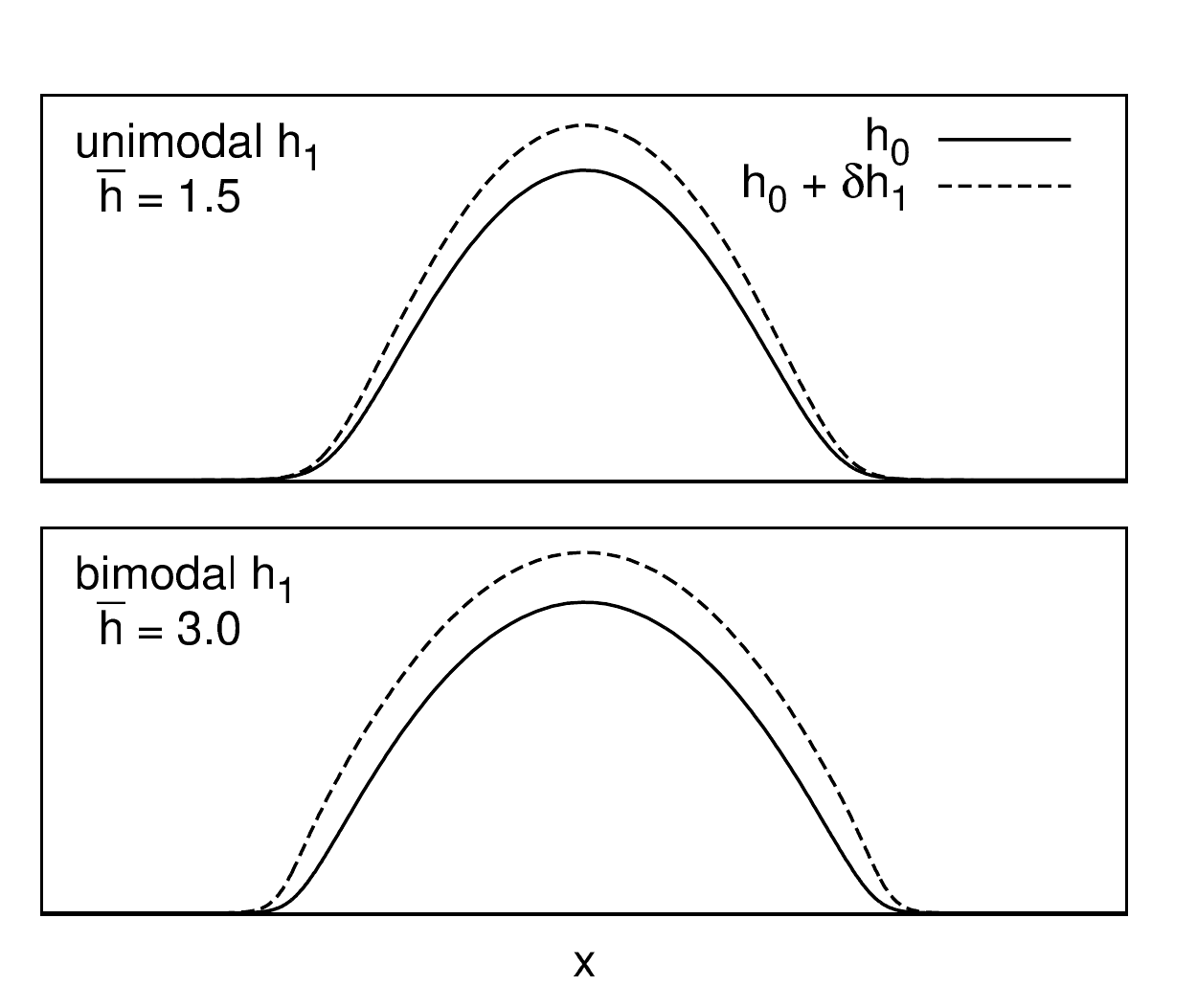}}
  \caption{Comparison of transversal instabilities with a unimodal eigenfunction $h_1(x)$ 
for small film heights (top, $\bar h=1.5$) and a bimodal $h_1(x)$ for larger film heights 
(bottom, $\bar h=3.0$). Shown are the steady ridge profiles (solid curves) together with the 
sum of the ridge profile and the eigenfunction multiplied by a small constant $\delta$ 
(dashed curve). In the first case the contact line stays fixed, in the second case it is shifted.} 
\label{fig:uni_bimodal}
\end{figure}

For $\bar h_1 < \bar h < \bar h_2$, the one-dimensional solutions have a pronounced droplet 
shape (cf.~solution for $\bar h=1.5$ in Fig.~\ref{fig:sol_trans}~a)). The corresponding 
two-dimensional ridge solutions are not stable, but this instability is not the Rayleigh-Plateau-like 
instability that leads to the formation of bulges. This can be inferred from the shape 
of the critical eigenfunction $h_1(x)$, which is depicted in Fig.~\ref{fig:sol_trans}~b) for 
different $\bar h$. For a Rayleigh-Plateau-like instability, $h_1(x)$ is supposed to have 
two symmetric maxima corresponding to the broadening of the ridge (cf.~Fig.~\ref{fig:uni_bimodal}).

In the region $\bar h_1 < \bar h < \bar h_2$, however, the function
$h_1(x)$ has only one maximum. Therefore this instability is related
to the formation of droplets \emph{on} the MWS and therefore belongs
to the growth regimes I to III observed in the experiments, which are
sketched in Fig.~\ref{fig:experiments2}.

In the region $\bar h_2 < \bar h < \bar h_3$ the critical
eigenfunction still has only one maximum, but the growth rates are
negative for finite wavenumbers (and zero for $q=0$).  In this region
a ridge on the MWS is stable because it is no longer possible for the
ridge to form droplets \textit{on} the MWS. This can be seen from the
critical eigenfunction at $\bar h=2.1$ in
Fig.~\ref{fig:sol_trans}~b). It becomes broader than the MWS. Since in
this region it is not yet energetically favorable for the ridge to
leave the stripe, the ridge stabilizes. This corresponds to the growth
regime V seen in the experiments.

For mean film heights above $\bar h_3$ it becomes energetically favorable for the liquid 
to cover also part of the LWS and the maximal growth rate becomes positive (cf.~Fig.~\ref{fig:beta_hmean} (a)).
At a critical film thickness of $\bar h^*\approx 2.5$, slightly greater than $\bar h_3\approx 2.4$, 
the critical eigenfunction undergoes a shape transition from unimodal to bimodal (see Fig.~\ref{fig:uni_bimodal}). 
The resulting instability is the Rayleigh-Plateau-like instability that leads to the formation of bulges. 
Therefore, the region $\bar h > \bar h_3$ corresponds to the growth regime VI (cf.~Fig.~\ref{fig:experiments2}).

For small wettability contrasts ($\rho \lesssim 0.43$) we observe the same shape transition 
in the critical eigenmode, but the maximal growth rate remains positive for $\bar h$ above 
the curve in Fig.~\ref{fig:phase_portrait2}. This is demonstrated in Fig.~\ref{fig:beta_hmean} 
(b) for $\rho=0.3$.

% \begin{figure}[htb]
% %% Plot file: ~/Dropbox/AUTO/MyRuns/LinHetDrop/1CH/TANH/Plot/pl_beta_hmean_rho03.plt
%   \rotatebox{270}{\includegraphics[height=.49\textwidth]{beta_hmean_rho03}}
%   \caption{Maximal growth rates $\beta_{\text{max}}$ and the corresponding 
%$q_{\text{max}}$ (see Eq.~\eqref{eq:def_betamax}) against $\bar h$ for fixed $\rho=0.3$.} \label{fig:beta_hmean_rho03}
% \end{figure}
As a next step we take a look at the influence of the sharpness of the wettability transition. 
The dotted line in Fig.~\ref{fig:phase_portrait2} shows the stability threshold in the 
$[\rho,\bar h]$ plane for $l_s = 0.001$, i.\,e. for a sharper wettability transition. 
In this case, stable ridges are possible for lower wettability contrasts $\rho$. On the 
other hand, for $\rho \gtrsim 0.57$, the onset of the Rayleigh-Plateau instability is 
shifted towards lower $\bar h$. This is probably due to the fact that the effective 
width of the MWS decreases with decreasing $l_s$. Further decreasing $l_s$ does not 
lead to a significant change of the diagram, therefore $l_s = 0.001$ can already be 
seen as the limiting case of a true step function.

Finally, we remark that the instability threshold in
  Fig.~\ref{fig:phase_portrait2} reaches the value $\bar h \approx
  1.26$ at $\rho=0$ as is the case for the sinusoidal wettability
  modulation in Fig.~\ref{fig:phase_portrait} above. This value of
$\bar h \approx 1.26$ corresponds, as it should be, to the
analytically obtained threshold of the spinodal instability of a flat
film, which is given by $\partial \varPi/\partial h=0$.  This
reinforces the interpretation that the first instability for small
$\rho$ and $\bar h$ is a spinodal instability. As one increases $\rho$
from zero to a small finite values while keeping $\bar h$ constant,
the one-dimensional flat film solutions change towards more and more
pronounced droplet solutions with increasing maximal film
heights. This explains why the instability threshold with respect to
$\bar h$ decreases with increasing $\rho$.

\subsection{Comparison of diffusive and convective transport}
\label{sec:trans}

\begin{figure}[h!tb]
\includegraphics[height=.49\textwidth]{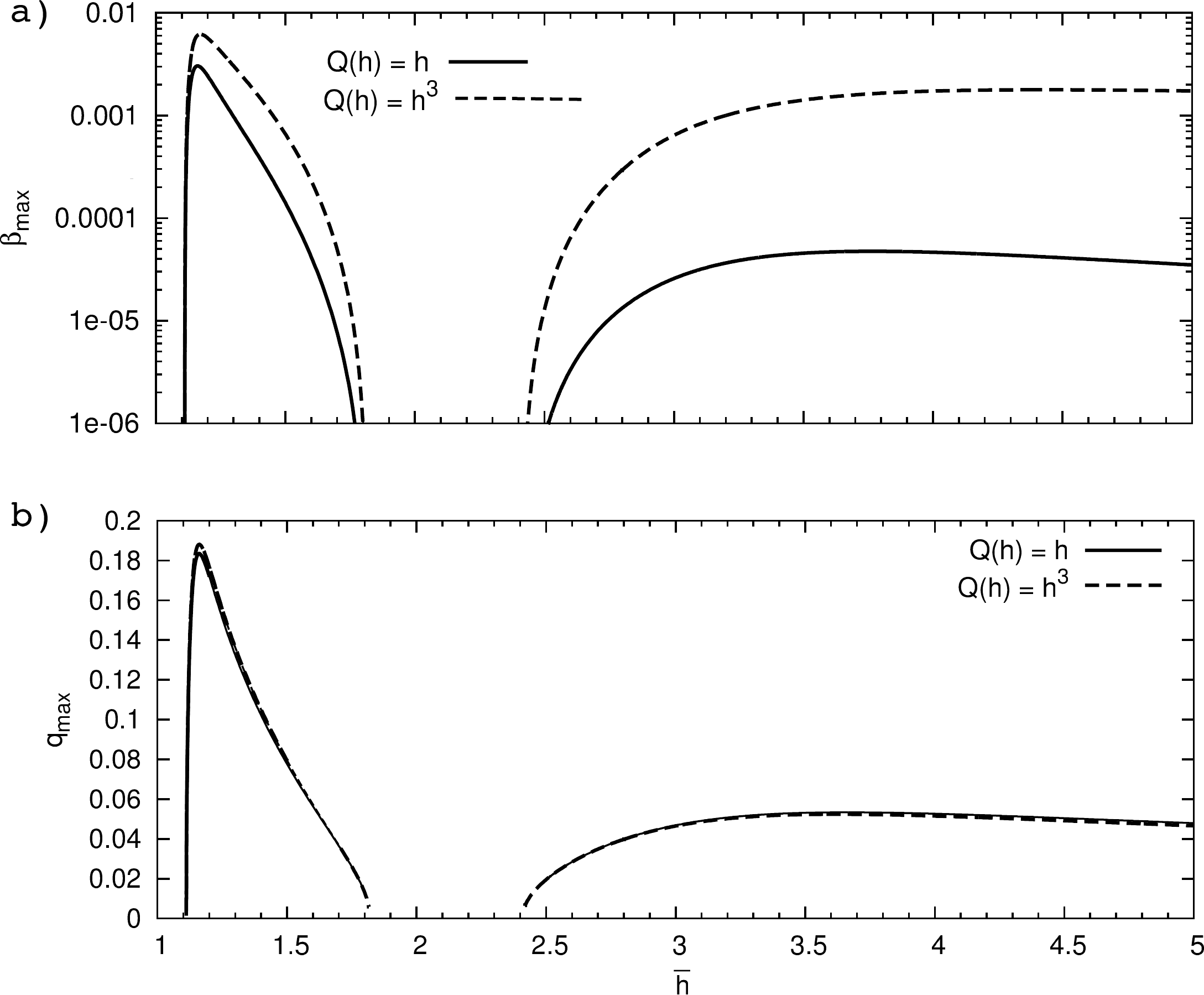}
\caption{a) Maximal growth rates of the transversal instability for $\rho=0.5$ 
against $\bar h$. The dotted line corresponds to a convective mobility $Q(h)=h^3$. This curve 
is the same as in Fig.~\ref{fig:beta_hmean}~a), but in logarithmic scaling. The 
solid line corresponds to a diffusive mobility term $h$. b) The corresponding fastest growing wavenumbers. The growth rates are 
significantly larger for the convective instability, especially in the bulge 
formation regime. The fastest growing wavenumbers are almost identical for both 
mobilities.} \label{fig:compare_beta_hmean}
\end{figure}

Equation \eqref{eq:tfe_stat}, which determines the equilibrium morphologies, does not depend 
on the mobility term $Q(h)$ as long as it is non-zero, whereas in the eigenvalue equation 
\eqref{eq:eval} only $Q(h_0)$ enters. Hence, the mobility has an effect on the 
absolute values of $\beta$ when $\beta \neq 0$, but does not influence the position 
of the instability thresholds $\beta=0$ in parameter space. 
This is demonstrated in Fig.~\ref{fig:compare_beta_hmean}, where we
plot, as in Fig.~\ref{fig:beta_hmean} (a), the maximum of the
dispersion relation $\beta_{max}$ against the mean film height $\bar
h$ for convective and diffusive mobility functions $Q(h)\sim
  h^3$ and $Q(h)\sim h$, respectively (note the semi-log scale in the
  upper panel). Although the absolute values are dramatically smaller
  for the diffusive mobility $Q(h)=h$, the stability thresholds
  $\beta_{max}=0$ remain at the same position, as can be seen more
  clearly in the bottom panel, where the corresponding $q_{max}$ is
  shown as a function of $\bar h$.
%
% \begin{figure}[htb]
% \begin{tabular}{l}
%  a)\\
%  %% Plot file: ~/AUTO07p/MyRuns/LinHetDrop/1CH/TANH/linear_mob/Plot/pl_cmp_bhmean_paper.plt
% %   \rotatebox{270}{\includegraphics[height=.49\textwidth]{cmp_bhmean_paper}}
%   \rotatebox{270}{\includegraphics[height=.45\textwidth]{compare_beta_hmean}}\\
% b)\\
% %% Plot file: ~/AUTO07p/MyRuns/LinHetDrop/1CH/TANH/linear_mob/Plot/pl_compare_qmax_hmean.plt
%   \rotatebox{270}{\includegraphics[height=.45\textwidth]{compare_qmax_hmean}}
% \end{tabular}
%   \caption{a) Maximal growth rates of the transversal instability for $\rho=0.5$ 
% against $\bar h$. The dotted line corresponds to a convective mobility \ttsveta{$Q(h)=h^3$}. This curve 
% is the same as in Fig.~\ref{fig:beta_hmean}~a), but in logarithmic scaling. The 
% solid line corresponds to a diffusive mobility term \ttsveta{$h$}. b) The corresponding fastest growing wavenumbers. The growth rates are 
% significantly larger for the convective instability, especially in the bulge 
% formation regime. The fastest growing wavenumbers are almost identical for both 
% mobilities.} \label{fig:compare_beta_hmean}
% \end{figure}
Therefore, we argue that the specific mobility that results from the
particular dominant transport process(es) does neither influence the 
linear stability of the ridges nor the observed equilibrium morphologies. Only the 
time scales at which the instabilities grow depend on the
mobility. This point will be further elucidated in the subsequent
section where we consider the time evolution in the fully nonlinear regime.

\section{Time simulations}
\label{sec:time}

In the previous section we have shown that there exist two
  different types of transversal instabilities that can occur in
  different ranges of values of the wettability contrast $\rho$:
  Inspection of the respective eigenfunctions indicates that for small
  volume a ridge decays into droplets that only cover the MWS, whereas
  for large volumes bulges evolve that also cover part of the
  LWS. Depending on the value of $\rho$ the two instabilities can be
  separated by a range of stable ridges. In the present section we
  test whether this finding of the linear analysis holds in the fully
  nonliner regime by performing direct numerical simulations of
  Eq.~\eqref{eq:tfe} in the two parameter regimes.

The numerical simulations are performed employing the alternating
direction implicit (ADI) method. Thereby the spatial and time
derivatives are discretised using second-order finite
differences. This results in a nonlinear system of equations that is
solved at each time step using Newton's method. In each Newton
iteration a linear system (Jacobian matrix of the nonlinear system) is
efficiently solved employing the ADI method ( see
Ref.~\citenum{Witelski2003, LiKF2012pf} for more
details). %In all simulations we set $\rho=0.5$, $l_s=0.03$, $x_A=0.3$, and $L_{per}=50$. The
%remaining parameters are given below. 

The first simulation is performed at low volume, for $\bar h=1.16$. This value corresponds to the 
maximum of $\beta_{max}$ that coincides to the maximum of $q_{max}$ (cf.~Fig.~\ref{fig:beta_hmean} a)), in order
to minimize the necessary computation time and domain size. The latter
is then selected as $50$ in $x$- and $100$ in $y$-direction. 

\begin{figure}[htb]
 \includegraphics[height=.3\textwidth]{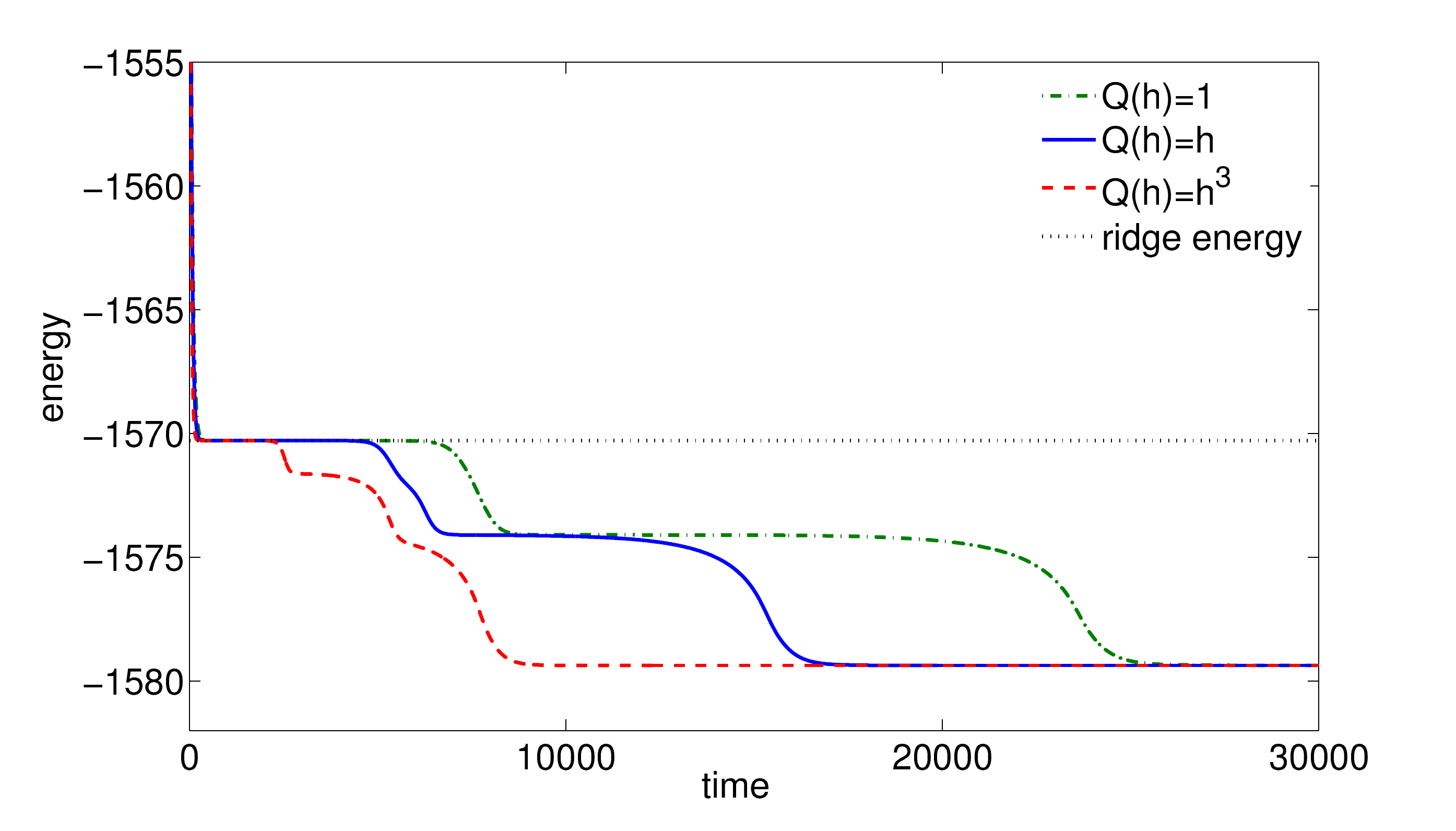}
 \caption{Time series of the energy during a direct numerical
   simulation of the thin film equation~\eqref{eq:tfe} with $\bar
   h=1.16$ for transport by surface diffusion (mobility $Q(h)\sim 1$,
   dash-dotted green curve), by bulk diffusion ($Q(h)\sim h$, solid blue line), and by convection ($Q(h)\sim h^3$, dashed red
   line). The configurations that correspond to the four energy
   plateaus for $Q(h)=h^3$ are plotted in
   Fig.~\ref{fig:snaps}.} \label{fig:energy_drop}
\end{figure}

\begin{figure}[htb]
 \begin{tabular}{ll}
  a) $t=10^3$ & b) $t=3\cdot 10^3$ \\
   \includegraphics[width=0.49\textwidth]{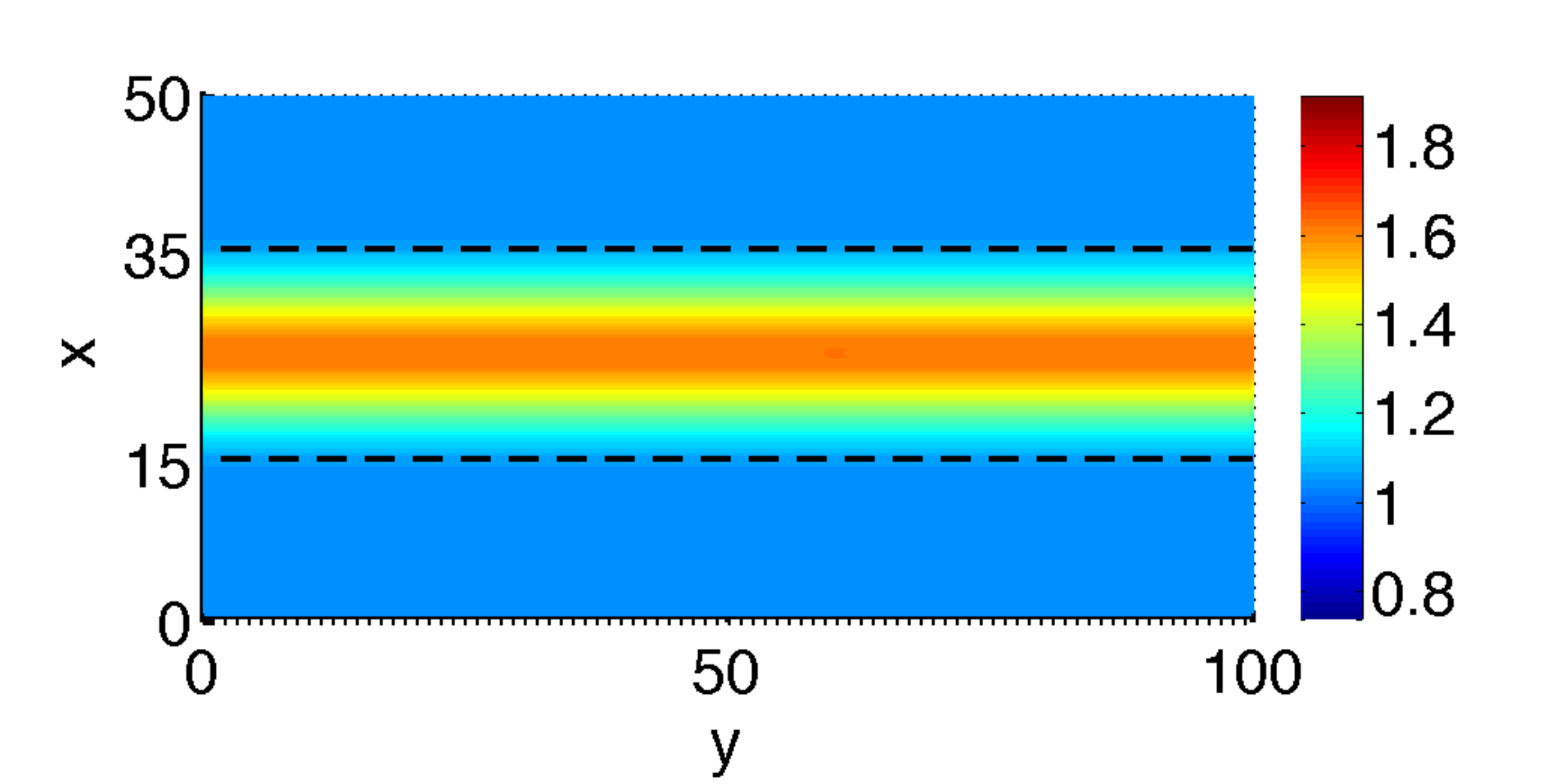} &  \includegraphics[width=0.49\textwidth]{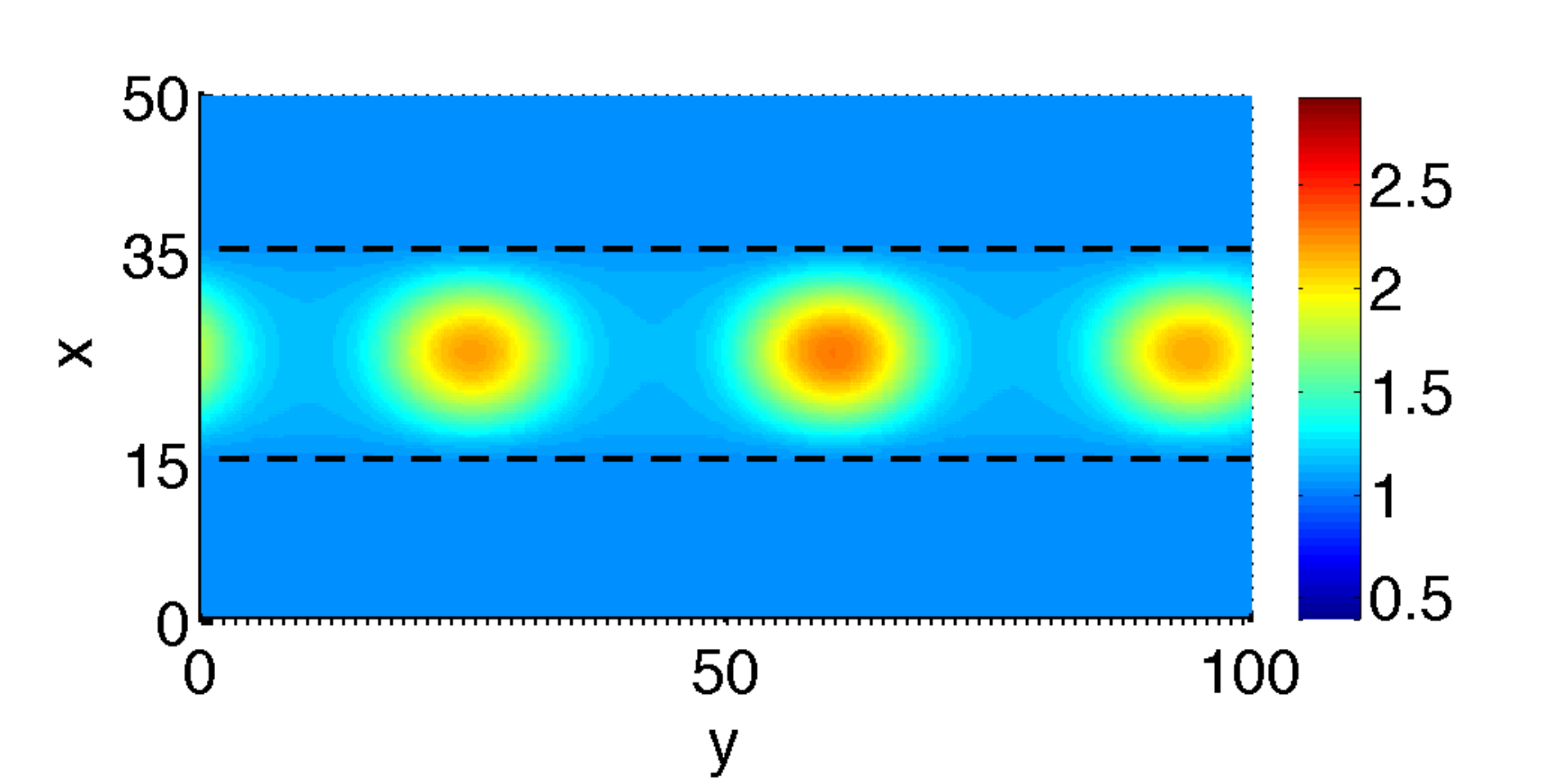}\\
   c) $t=6\cdot 10^3$ & d) $t=10^4$\\
   \includegraphics[width=0.49\textwidth]{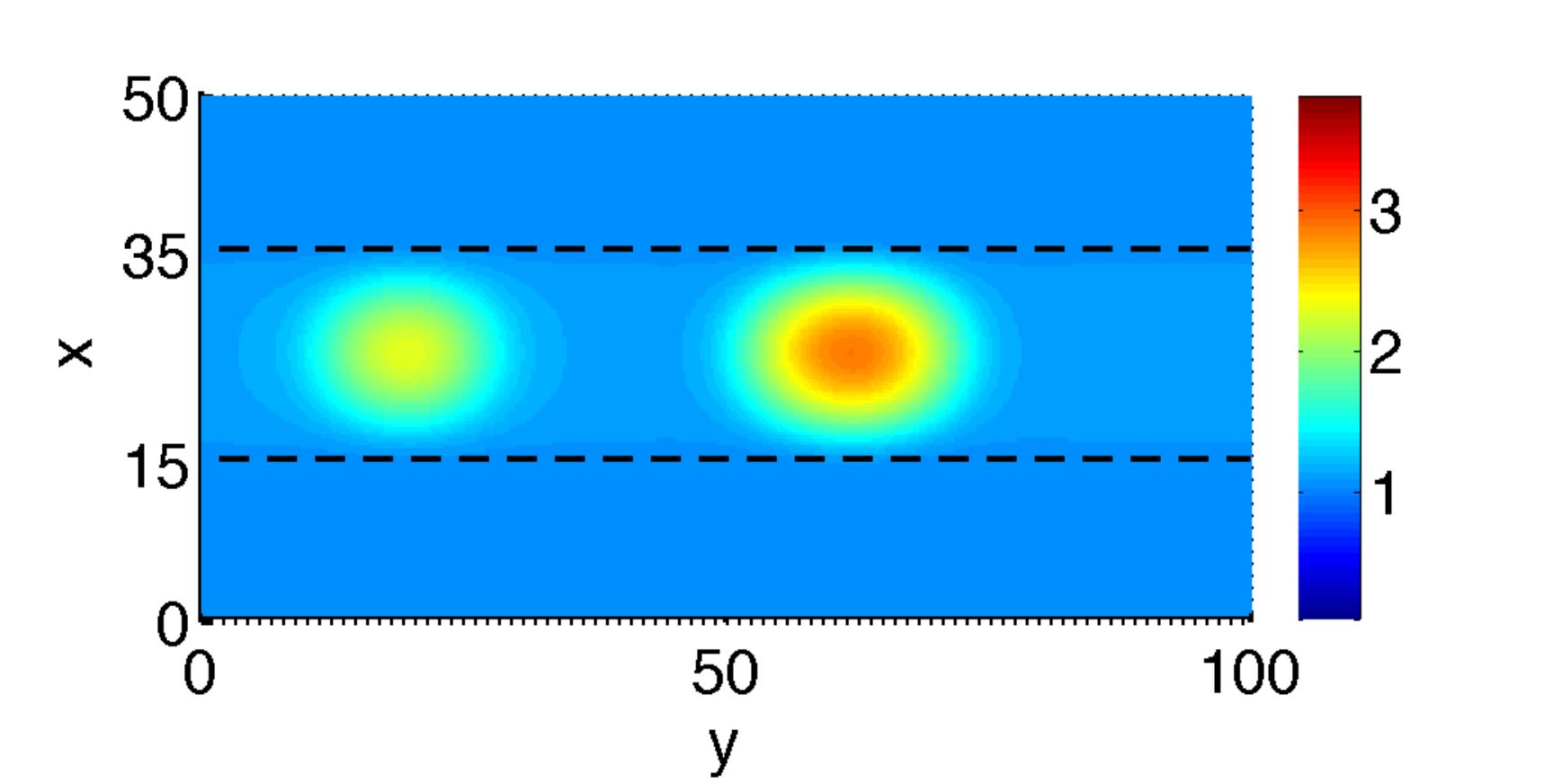} & \includegraphics[width=0.49\textwidth]{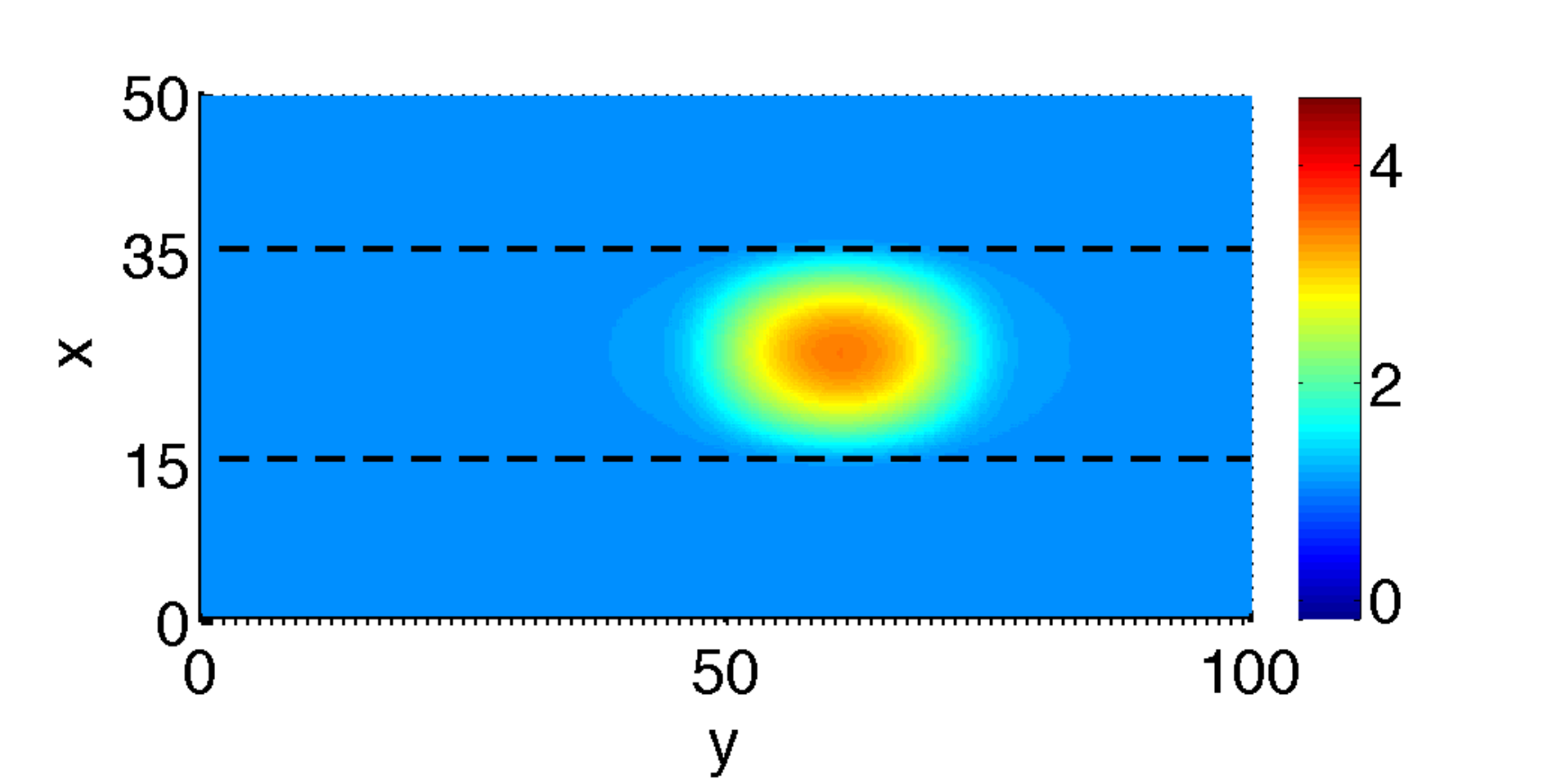}
 \end{tabular}    
 \caption{Snapshots of a direct time simulation of the thin film
   equation~(\ref{eq:tfe}) with $\bar h=1.16$, $l_s=0.03$, $x_A=0.3$,
   $L_{per}=50$ in the case of purely convective transport (mobility
   $Q(h)= h^3$).  The borders of the MWS is
   indicated by dashed (black) lines.  Figure~\ref{fig:energy_drop}
   shows the corresponding time series of the energy $F(t)$.
   The four snapshots at times $t=1000,\, 3000,\, 6000,\, 10000$
   correspond to the four energy plateaus in
   Fig.~\ref{fig:energy_drop}.}
 \label{fig:snaps}
\end{figure}

\noindent In the course of the time evolution we measure the non-dimensional free energy \\
$F(t)=\int \md x\,\md y \l[ \tfrac12 (\nabla h)^2 + f(h,\,x) \r]$, where $f(h,\,x)$ now stands for the
  non-dimensional wetting potential. Note that the total time
derivative of the free energy is given by $\dfrac{d F}{d t}=-\int
Q(h)\,\left(\dfrac{\delta F}{\delta h}\right)^2\, \md x\,\md y\,.$ As
the mobility function $Q(h)$ is always positive, the final expression
is negative, i.e., $F$ is a Lyapunov functional. As a consequence, linearly stable and unstable steady solutions of the
  thin film equation correspond to local minima or saddle points of
  the free energy functional $F$, respectively. The resulting
dependency of energy on time is shown in Fig.~\ref{fig:energy_drop}.
%
% \begin{figure}[htb]
%   \rotatebox{270}{\includegraphics[height=.49\textwidth]{energy_h116}}
%  \caption{Time series of the energy during a simulation of the thin film equation with $\bar h=1.16$ employing a convective (dashed curve) and diffusive (solid curve) mobility. The configurations that correspond to the four energy plateaus are plotted in Fig.~\ref{fig:snaps}.} \label{fig:energy_drop}
% \end{figure}
In this energy time series obtained with the convective
  mobility function $Q(h)$ (dashed red curve) one can identify four
  plateaus of different length where the evolution approaches
  different steady states of the system.  The first three correspond
  to unstable steady profiles, while the final one corresponds to the
  absolutely stable steady state configuration. All four
configurations for $Q(h)=h^3$ are depicted in Fig.~\ref{fig:snaps}.
%
% \begin{figure}[htb]
% %% Plot file: /local1/c_honi01/ThinFilm/2D/FD/SPARSE/TANH/o6/Run01/pl_cut_ridge.plt
%   \rotatebox{270}{\includegraphics[height=.45\textwidth]{cut_ridge}}
%   \caption{Comparison between the time simulation and the one-dimensional continuation. The solid line 
% shows the result of the one-dimensional  continuation, the dots depict a cut through the ridge configuration 
% at $y=0$ and $t=1000$. \sout{The agreement is perfect.}} \label{fig:comp_ridge}
% \end{figure}
%

\begin{figure}[htb]
%% Plot file: /local1/c_honi01/ThinFilm/2D/FD/SPARSE/TANH/o6/Run01/pl_cut_ridge.plt
  \rotatebox{270}{\includegraphics[height=.45\textwidth]{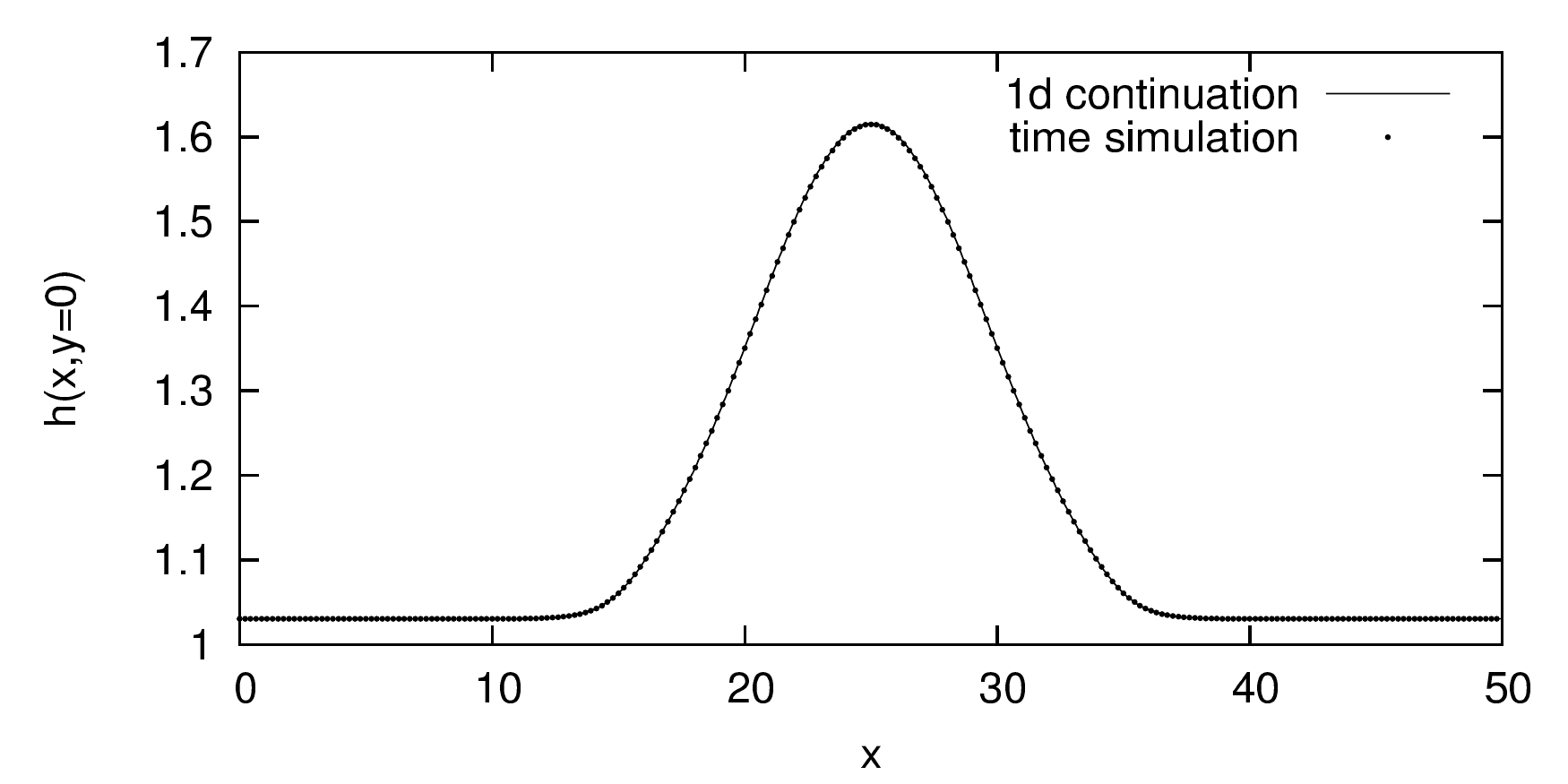}}
  \caption{Comparison between solutions that result from the
    one-dimensional continuation (solid line) and from the direct time
    simulation (where the dotted line gives the profile of a
    cross-section at $y=0$ and $t=1000$ of the transversally invariant
    ridge state assumed for a long intervall in the time evolution,
    cf. Fig.~\ref{fig:snaps}~a)).  } \label{fig:comp_ridge}
\end{figure}

The first configuration, which is assumed very quickly starting from a
flat film perturbed by small-amplitude noise, is a transversally
invariant ridge that is identical to the one obtained above through
one-dimensional steady state continuation. In Fig.~\ref{fig:snaps}~a)
and Fig.~\ref{fig:comp_ridge}, the solutions resulting from the direct
numerical simulation (dotted line) and of the one-dimensional
continuation (solid line) are compared.
The fact that this unstable steady state is assumed for a considerably
long time during the time simulation retrospectively justifies the
approach of the transversal linear stability analysis of the previous
section.
After a certain time the homogeneous ridge develops transversal
modulations that grow exponentially and evolve into the three
droplet state that corresponds to the second plateau and is depicted
in Fig.~\ref{fig:snaps}~b). Two of the drops shrink and vanish in
subsequent coarsening events until only one droplet remains
(Fig.~\ref{fig:snaps}~c) and d)). The borders between the LWS and MWS
in Fig.~\ref{fig:snaps} are indicated by dashed black lines. From this
one can see that the droplets are completely located on the MWS,  in
agreement with the experimental case (I) and with the discussion in the previous section.

\begin{figure}[htb]
%% Plot file: ./TeSheng/data/W50_h116_Y100_hh3/pl_growth.plt
\rotatebox{270}{\includegraphics[height=.45\textwidth]{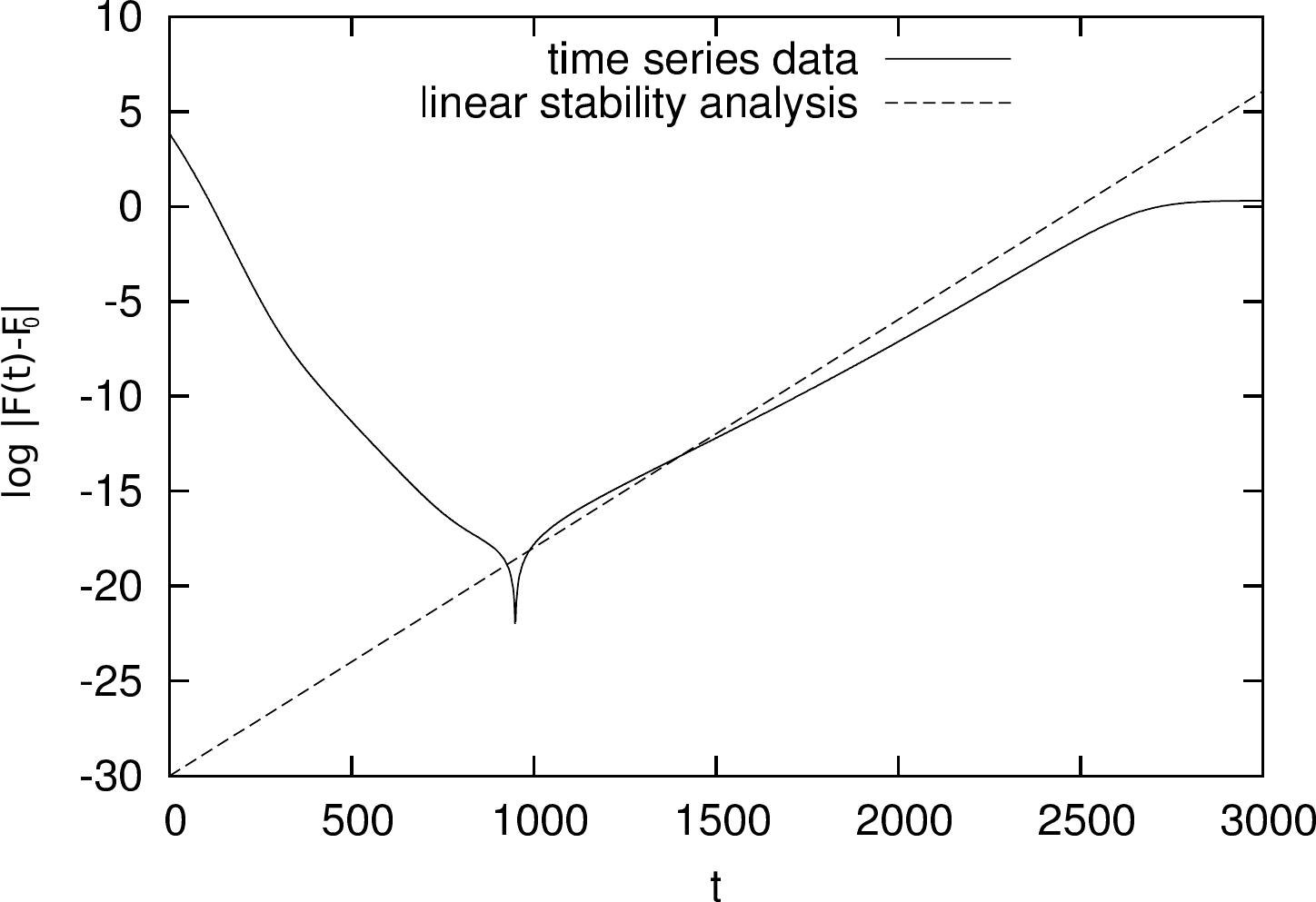}}
\caption{The solid line gives $\log |F(t) - F_0|$, where $F_0$ is the
  free energy of the steady transversally invariant ridge state $h_0$
  and $F(t)$ is the free energy at time $t$ during the time simulation
  in Fig.~\ref{fig:snaps}. The dashed curve shows the exponential
  growth with a growth rate $2 \beta_m$ that one expects from the
  linear stability analysis w.r.t.\ transversal modes.}\label{fig:engdiff_h116}
\end{figure}

To further characterize the departure of the time evolution
  from the ridge state of the first plateau,
  Fig.~\ref{fig:engdiff_h116} (solid line) shows the time evolution of
  the logarithm of the difference $\bigtriangleup F$ between the
  energy $F(t)$ and the reference energy $F_0$ of the ridge solution
  $h=h_0(\mathbf{x})$. One observes that for the time interval when
the system is close to the unstable ridge solution
(cf. Fig.~\ref{fig:snaps}~a) and b)), the energy $\bigtriangleup F$
grows exponentially with a growth rate $2\,\beta_{\mathrm{max}}$
(dashed curve) as expected from the transversal linear stability
analysis. Here, $\beta_{\mathrm{max}}=\beta(q_{\mathrm{max}})$ is the
growth rate, calculated for the most unstable mode of the
transversally invariant ridge solution. Indeed, an expansion of the
energy $F(t)$ about the ridge solution $h_0$ with the perturbed
solution of the form
$h=h_0+\varepsilon\,h_1(\mathbf{x})\,\exp(\beta\,t)$ reads
\begin{eqnarray*}
 F(t)&=&F_0+\varepsilon\,\int\,d \mathbf{x}\,\frac{\delta F}{\delta h}\biggl|_{h=h_0}h_1(\mathbf{x})\,e^{\beta\,t}\\
 &&+\varepsilon^2\,\int\,d \mathbf{x}\,\frac{\delta^2 F}{\delta h^2}\biggl|_{h=h_0}h_1(\mathbf{x})^2\,e^{2\,\beta\,t}+\mathrm{h.\,o.\,t.}\,.
\end{eqnarray*}

As the linear part in $\varepsilon$ of this expansion vanishes at $h=h_0$, the nontrivial term of lowest order is the quadratic one, explaining the growth rate of $2\,\beta$.

Note that the transient behaviour of the energy $F(t)$ changes when
employing different mobility functions $Q(h)$. The solid blue and
green dash-dotted lines in Fig.~\ref{fig:energy_drop}) give the result for
linear (bulk diffusion) and constant (surface diffusion) mobilities,
respectively. Remarkably, one clearly sees that the same energy
plateaus appear independent of the chosen mobility, however, their
relative duration depends on the particular chosen transport
behaviour. For instance, with the convective mobility after the ridge
configuration, the evolution visits the three drop solution that is
still visible as a shoulder with the linear mobility and is only
barely visible (when zooming in) with the constant mobility. The
two drop solution is clearly visited by all evolutions and the final
configuration (corresponding to the single drop solution) 
is the same for all mobilities $Q(h)$ what is expected as $F$ is a
Lyapunov functional, independent of the mobility.

%
% \begin{figure}[htb]
% %% Plot file: ~/AUTO07p/MyRuns/LinHetDrop/1CH/TANH/CalcGrowthRates/pl_disp_rel.plt
% \rotatebox{270}{\includegraphics[height=.45\textwidth]{disp_rel_h116}}
% \caption{Dispersion relation for the unstable mode of a stationary ridge for a system 
% with $l_s=0.03$, $x_A=0.3$, $L_{per}=50$, $\bar h=1.16$ and $L_x=50$ with a convective mobility $Q(h) = h^3$. The corresponding 
% time simulation has $L_y=100$. The allowed mode with maximal growth rate is
% for $q_m=6\pi/100\approx 0.1885$. The corresponding growth rate is 
% $\beta_m=6.0125\cdot10^{-3}$.} \label{fig:disp_rel_h116}
% \end{figure}
% \ttsveta{
% $$
% F(t)=F(h_0)+\varepsilon\,\int\,d \mathbf{x}\,\frac{\delta F}{\delta h}\biggl|_{h=h_0}h_1(\mathbf{x})\,e^{\beta\,t}+\varepsilon^2\,\int\,d \mathbf{x}\,\frac{\delta^2 F}{\delta h^2}\biggl|_{h=h_0}h_1(\mathbf{x})^2\,e^{2\,\beta\,t}+\mathrm{h.\,o.\,t.}\,.
% $$
% }
%In \ref{fig:w100} we perform the time simulation for $\bar{h} = 1.9$, $l_s=0.015$, $\rho=0.5$, $x_A=0.4$ and $L_{per}=100$. 
\begin{figure}[htb]
 \begin{tabular}{ll}
  a) $t=10^4$ & b) $t=2\cdot 10^4$\\
    \includegraphics[width=0.49\textwidth]{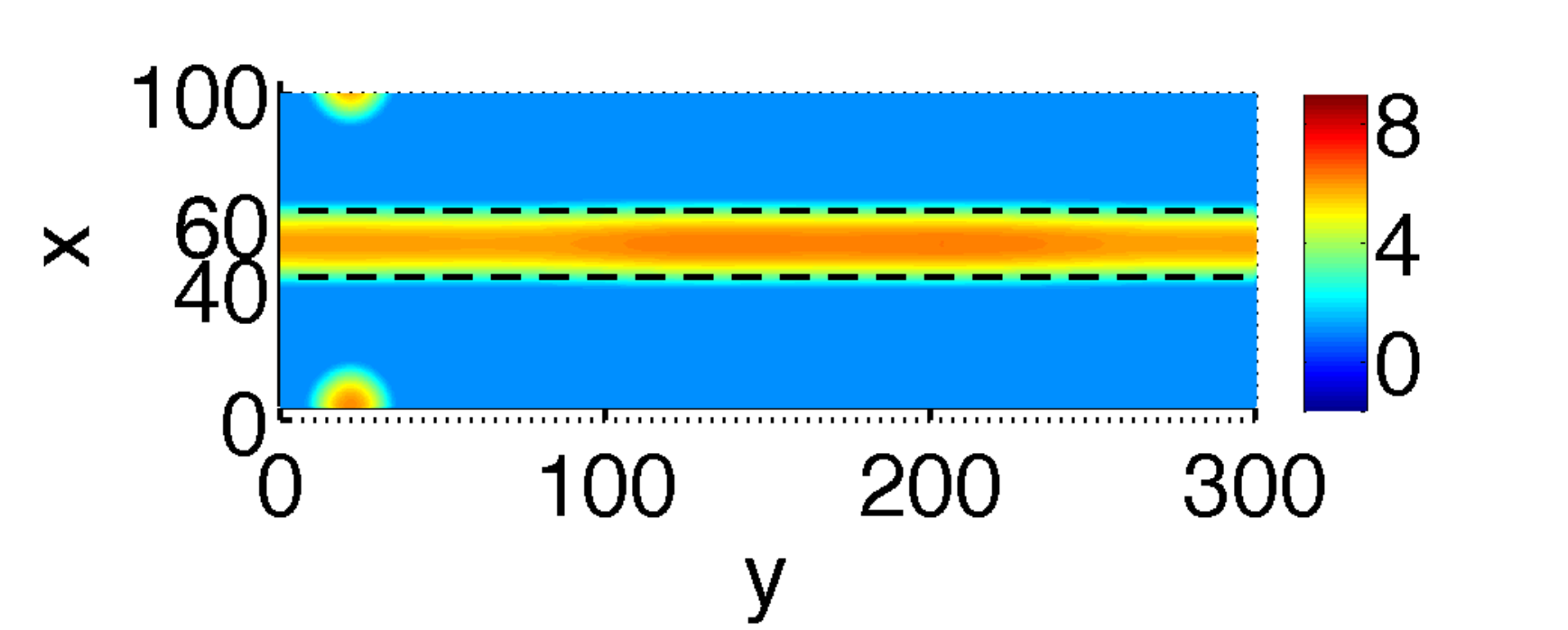} &  \includegraphics[width=0.49\textwidth]{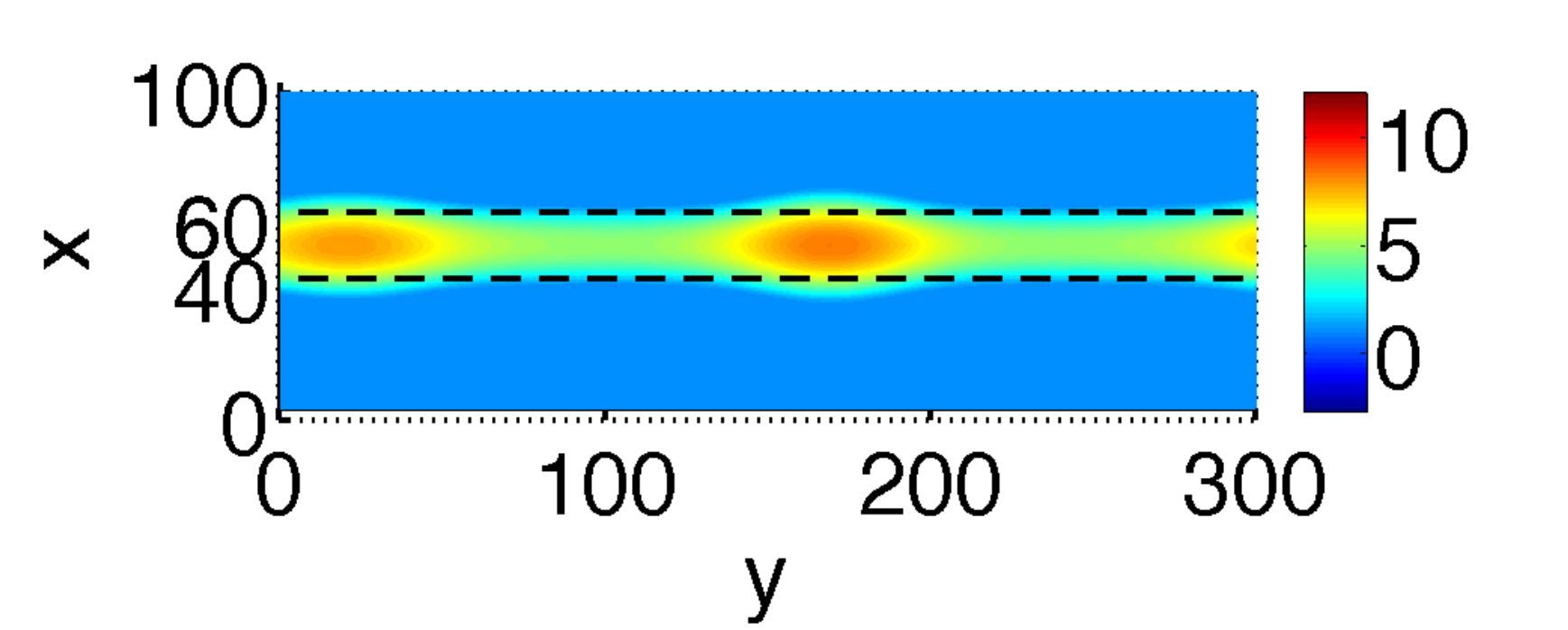}\\
   c) $t=3\cdot 10^4$ & d) $t=8\cdot 10^5$\\
    \includegraphics[width=0.49\textwidth]{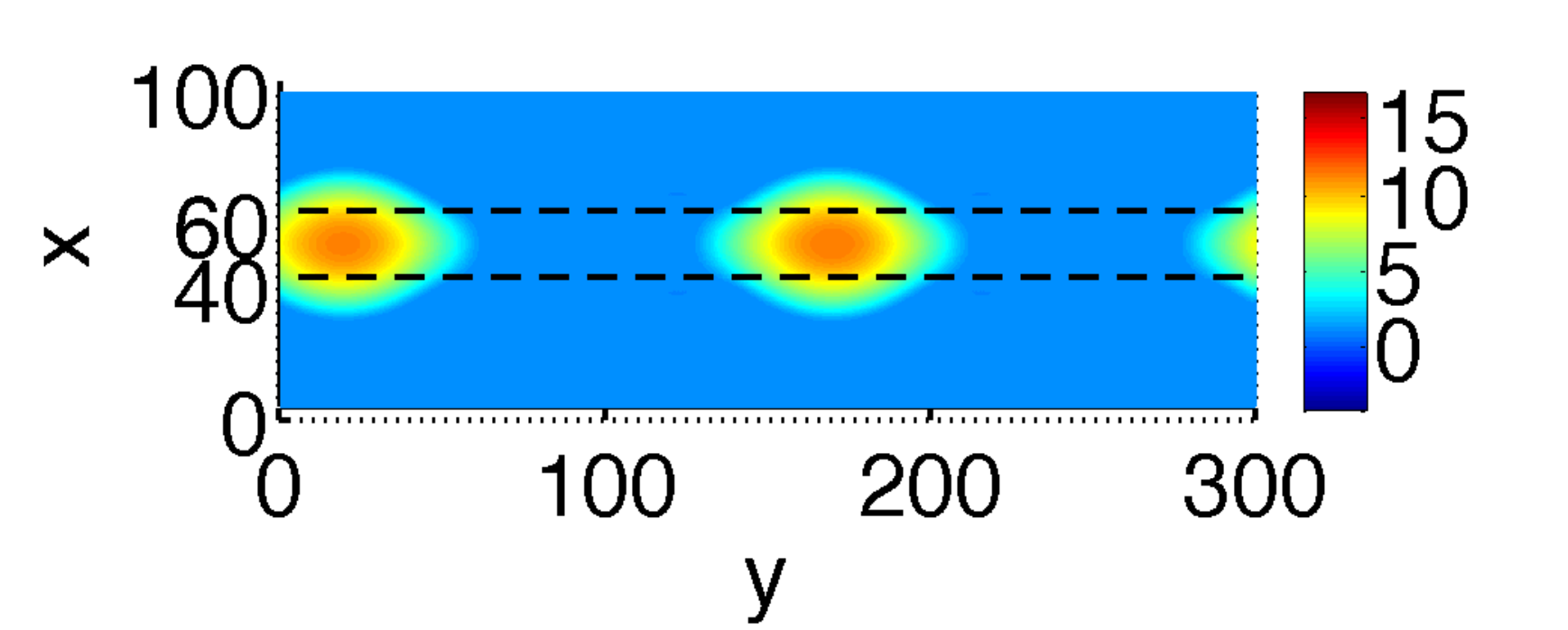}&\includegraphics[width=0.49\textwidth]{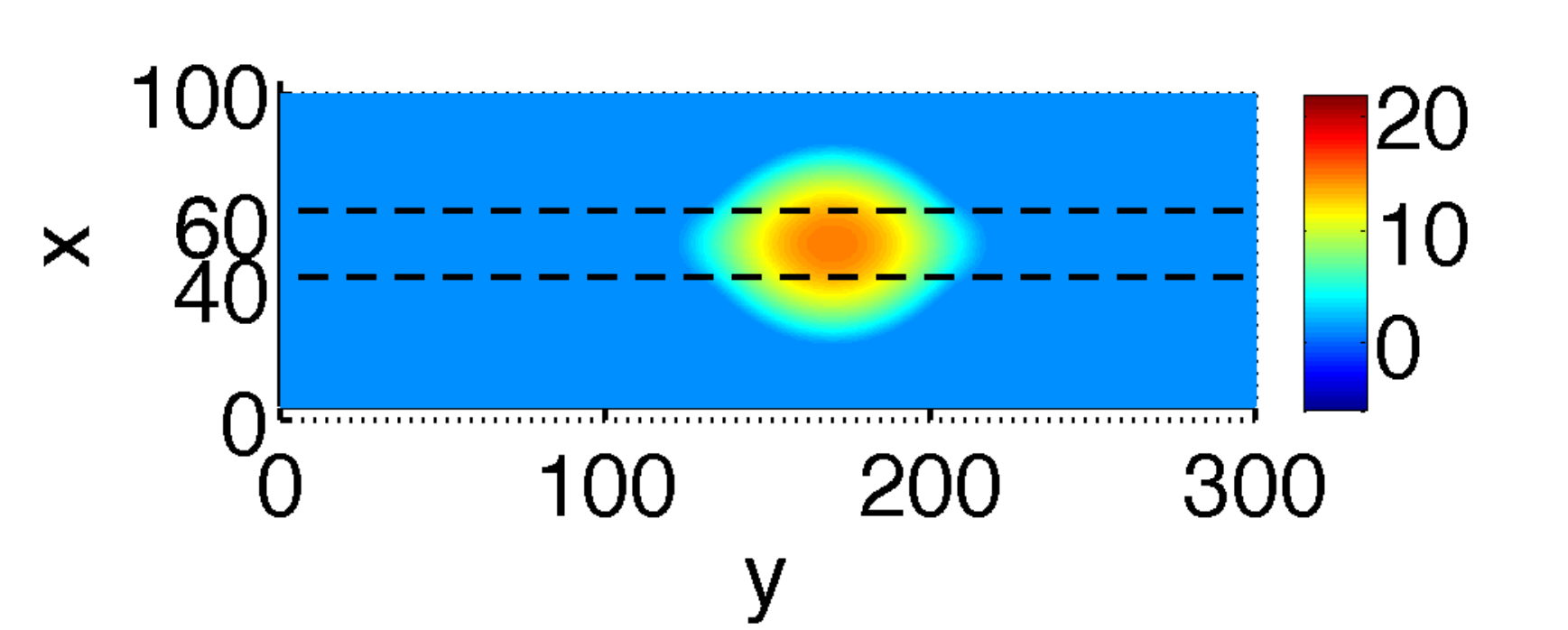}\\   
 \end{tabular}
 \caption{Snapshots of a direct time simulation of Eq.~(\ref{eq:tfe})
   at $h = 1.9$, $\rho= 0.5$, $l_s=0.015$, $x_A=0.4$ and $L_{per}=100$
   and a convective mobility $Q(h) = h^3$. Panels (a), (b), (c), and
   (d) show snapshots at times $t=10^4$, $t=2\cdot 10^4$, $t=3\cdot
   10^4$ and $t=8\cdot 10^5$, respectively. The borders of the more
   wettable stripes (MWS) is indicated by dashed (black) lines. }
\label{fig:w100}
\end{figure}

\begin{figure}[htb]
 \begin{center}
    \includegraphics[width=0.49\textwidth]{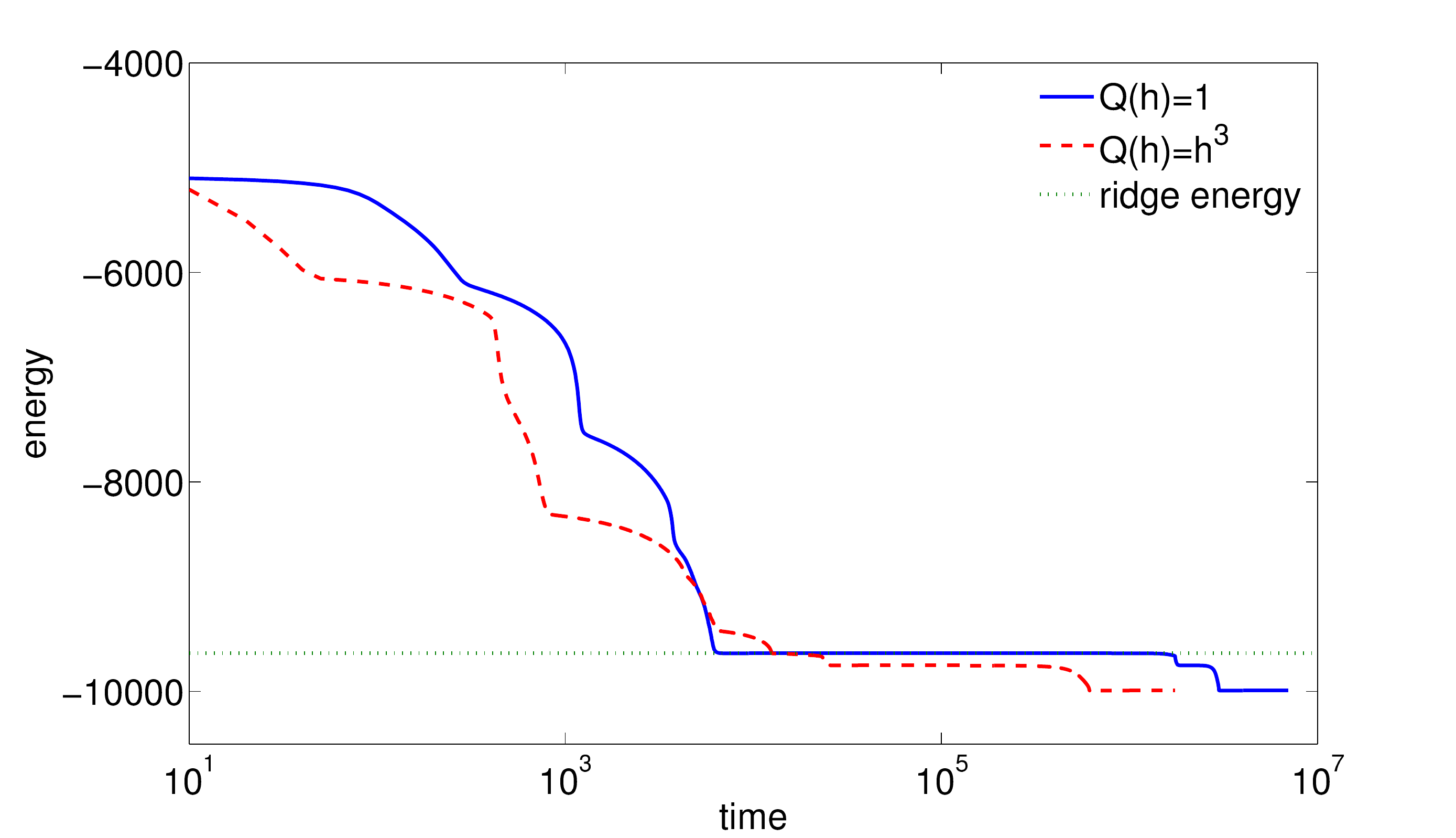}
 \end{center}
 \caption{Time series of the free energy during the time simulation of
   Eq~\eqref{eq:tfe} with $\bar h=1.9$ for transport by surface
   diffusion (mobility $Q(h)\sim 1$, solid blue line) and by
   convection ($Q(h)\sim h^3$, dashed red line). The configurations
   that correspond to last four energy plateaus for $Q(h)=h^3$ are
   given in Fig.~\ref{fig:w100}.}\label{fig:bulgeen}
\end{figure}

\begin{figure}[htb]
 \begin{center}
    \includegraphics[width=0.49\textwidth]{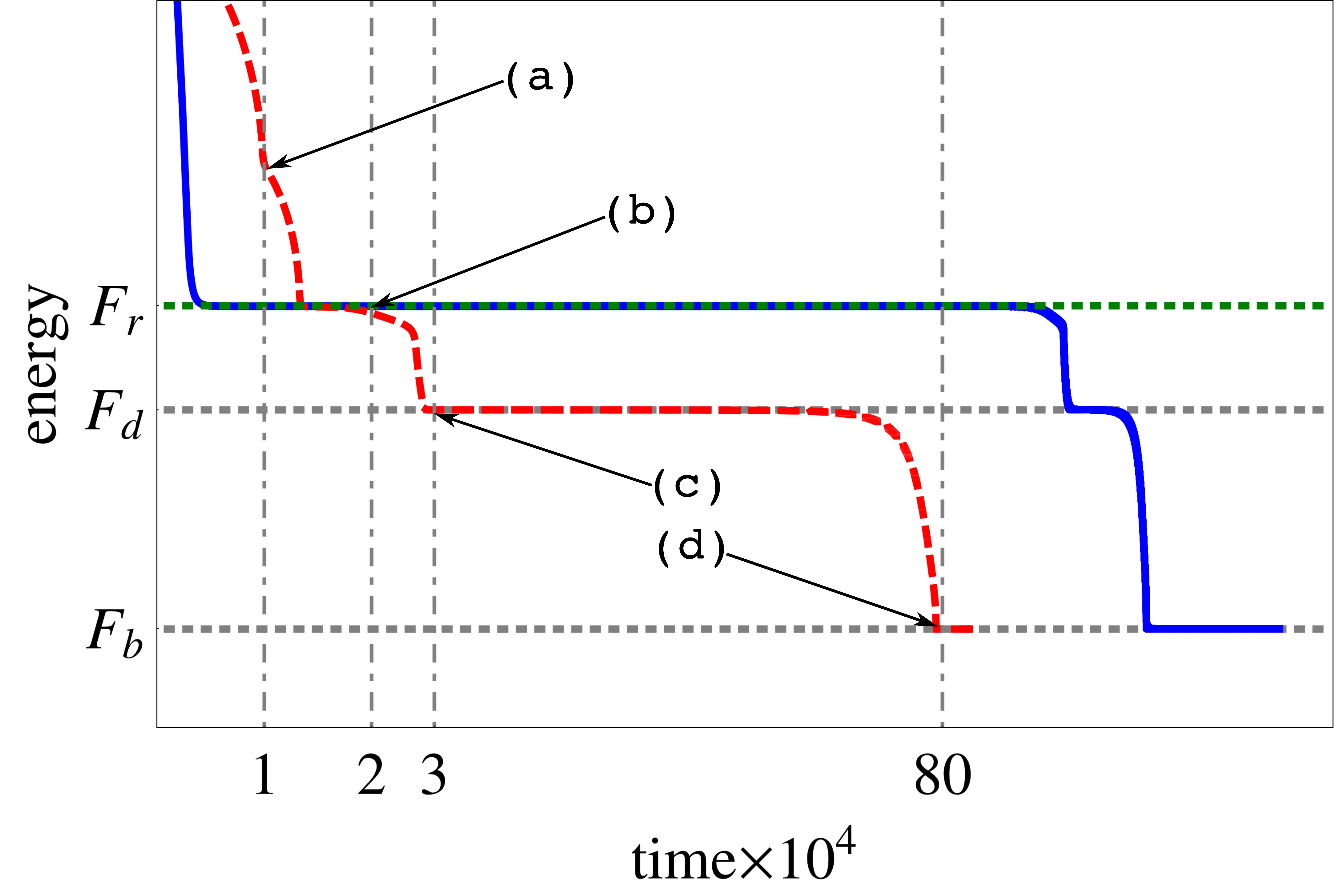}
 \end{center}
 \caption{A zoom into the final thousand time steps in the energy
   plot~(\ref{fig:bulgeen}) for the time simulation of
   Eq.~\eqref{eq:tfe} with $\bar h=1.9$ for transport by surface
   diffusion (mobility $Q(h)\sim 1$, solid blue line) and by
   convection ($Q(h)\sim h^3$, dashed red line). The letters (a)-(d) refer to snapshots from
   DNS (cf. Fig.~\ref{fig:w100}) that correspond to last four energy
   plateaus for $Q(h)=h^3$ (right).}\label{fig:zoombulge}
\end{figure}

Next, we analyse the time evolution in the parameter range where, according
to the linear stability analysis in the previous section, the second type of transversal instability should occur that is related to bulge formation.
The wettability contrast is kept at $\rho=0.5$, but a larger
initial film height $\bar{h}= 1.9$ is chosen. Figure~\ref{fig:w100} shows four
snapshots from this time simulation for a convective mobility $Q(h) =
h^3$, whereas Fig.~\ref{fig:bulgeen} gives the corresponding dependence
of the energy on time for convective ($\sim h^3$, red dashed line) and
surface diffusion ($\sim1$, solid blue line) mobilities $Q(h)$). In addition, Fig.~\ref{fig:zoombulge}
shows a zoom into the energy evolution for the final thousand time steps (in logarithmic scale). There, the (a)-(d) letters correspond to the
snapshots from the time simulations that correspond to the various plateaus.
As in the previous case, the transversally invariant ridge
is assumed quickly (Fig.~\ref{fig:w100}~a)) and forms the starting
point for the subsequent morphological transitions. Note, however, that
here the ridge solution is accompanied by a single small drop, situated on
the LWS. The existence of the state with this drop results in a shoulder in the
energy plot at about $t=10^4$ (cf. Fig.~\ref{fig:zoombulge}).  After a
short time, the drop shrinks and disappears and the system stays close
to the unstable steady transversally invariant ridge (with energy $F_r$) for some time (see the short
energy plateau in Fig.~\ref{fig:zoombulge} for $Q(h)=h^3$ close to
$F_r$). Eventually the ridge develops modulations in $y$-direction (see
panel b) of Fig.~\ref{fig:w100}). These modulations grow and form two
large bulges (Fig.~\ref{fig:w100}~c) and Fig.~\ref{fig:zoombulge} at
$F=F_d$). Finally, one of them vanishes in a coarsening event and only
one bulge remains (panel d) of Fig.~\ref{fig:w100} and
Fig.~\ref{fig:zoombulge} at $F=F_b$ ). From the dashed black line,
indicating the borders of wettability regime, one can see that in this
case the solution is not pinned to the MWS and also covers the
LWS. The emerged configurations resemble the experimentally  observed case VI 
(see Fig.~\ref{fig:experiments2}.)

\begin{figure}[ht!]
 \begin{tabular}{l}
  a) $t=10^4$\\
   \includegraphics[width=0.49\textwidth]{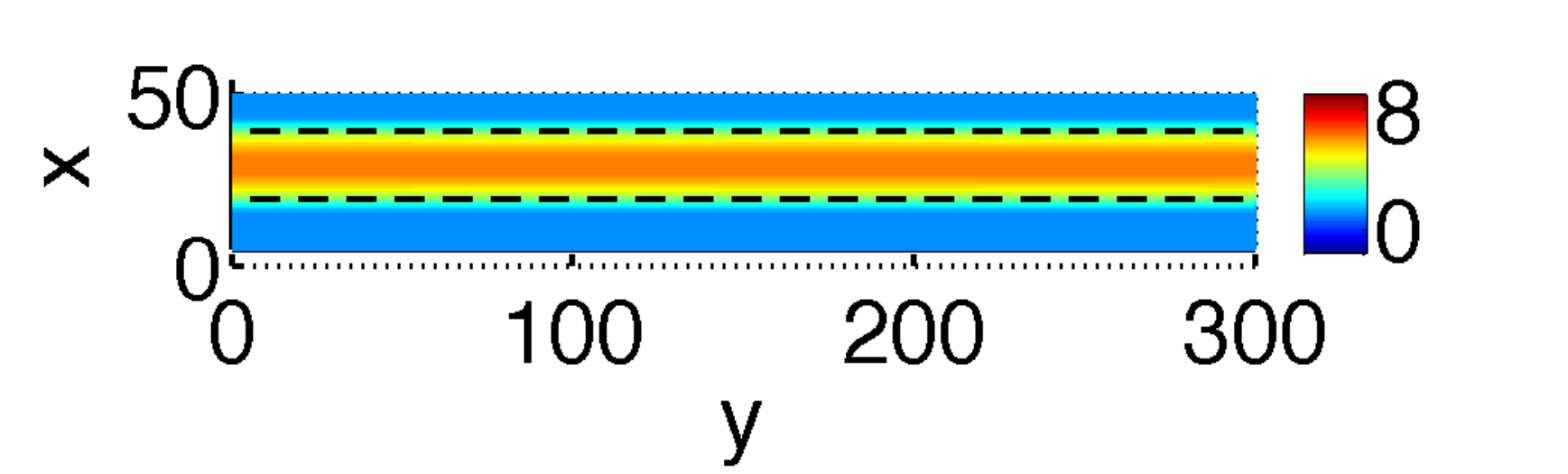}\\
   b) $t=5\cdot 10^4$\\
    \includegraphics[width=0.49\textwidth]{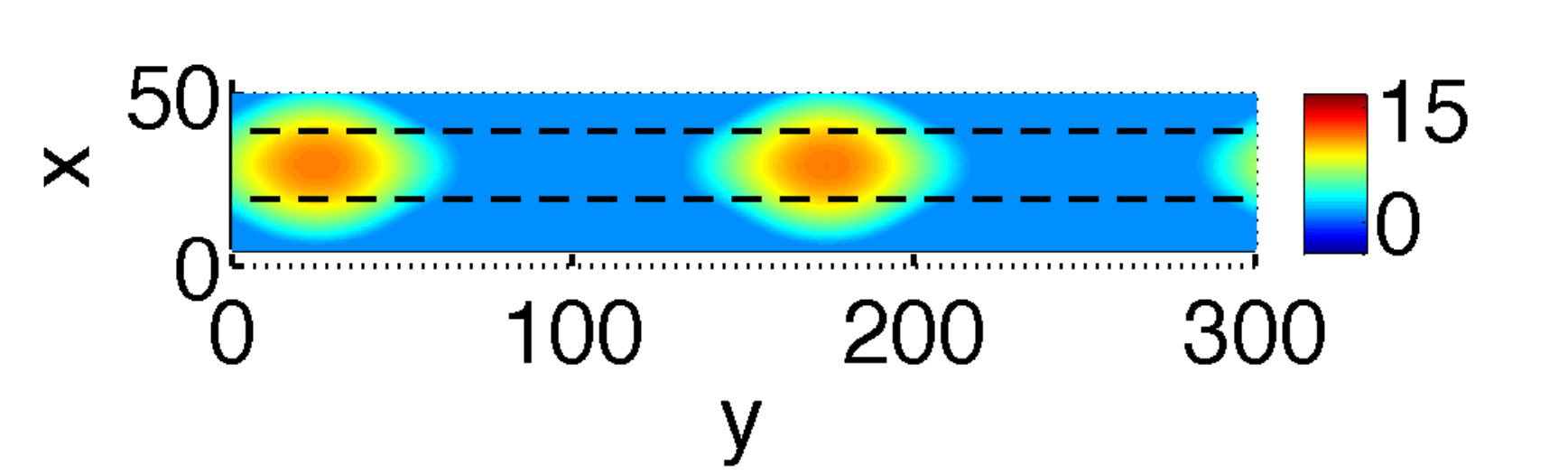}\\
    c) $t=1.5\cdot 10^6$\\
      \includegraphics[width=0.49\textwidth]{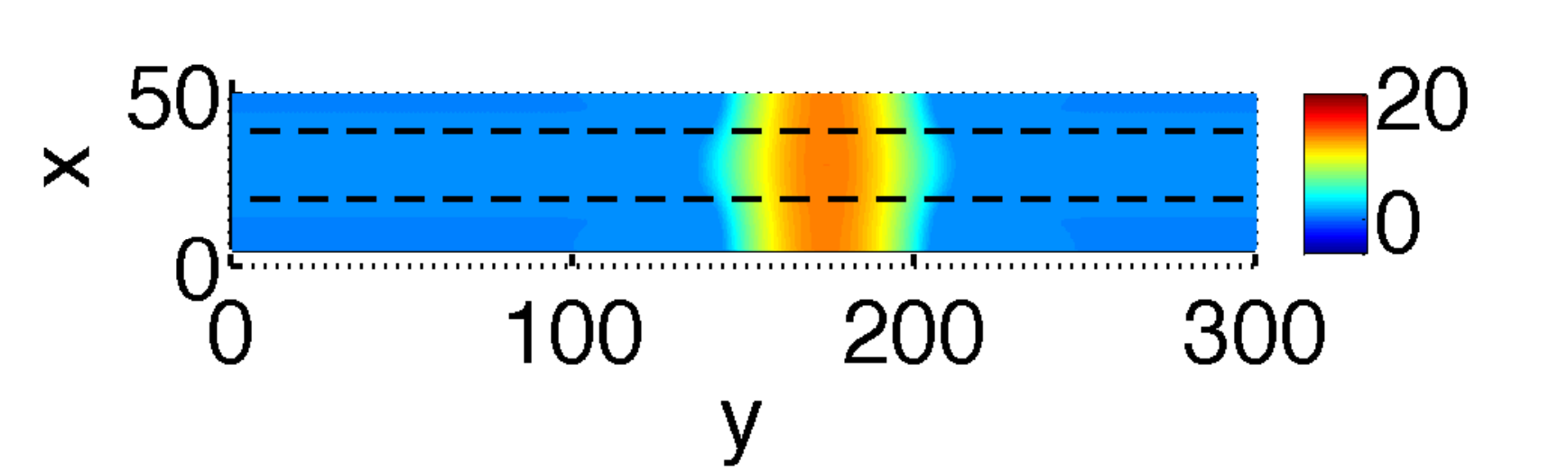}\\
    d) \\
     \includegraphics[width=0.45\textwidth]{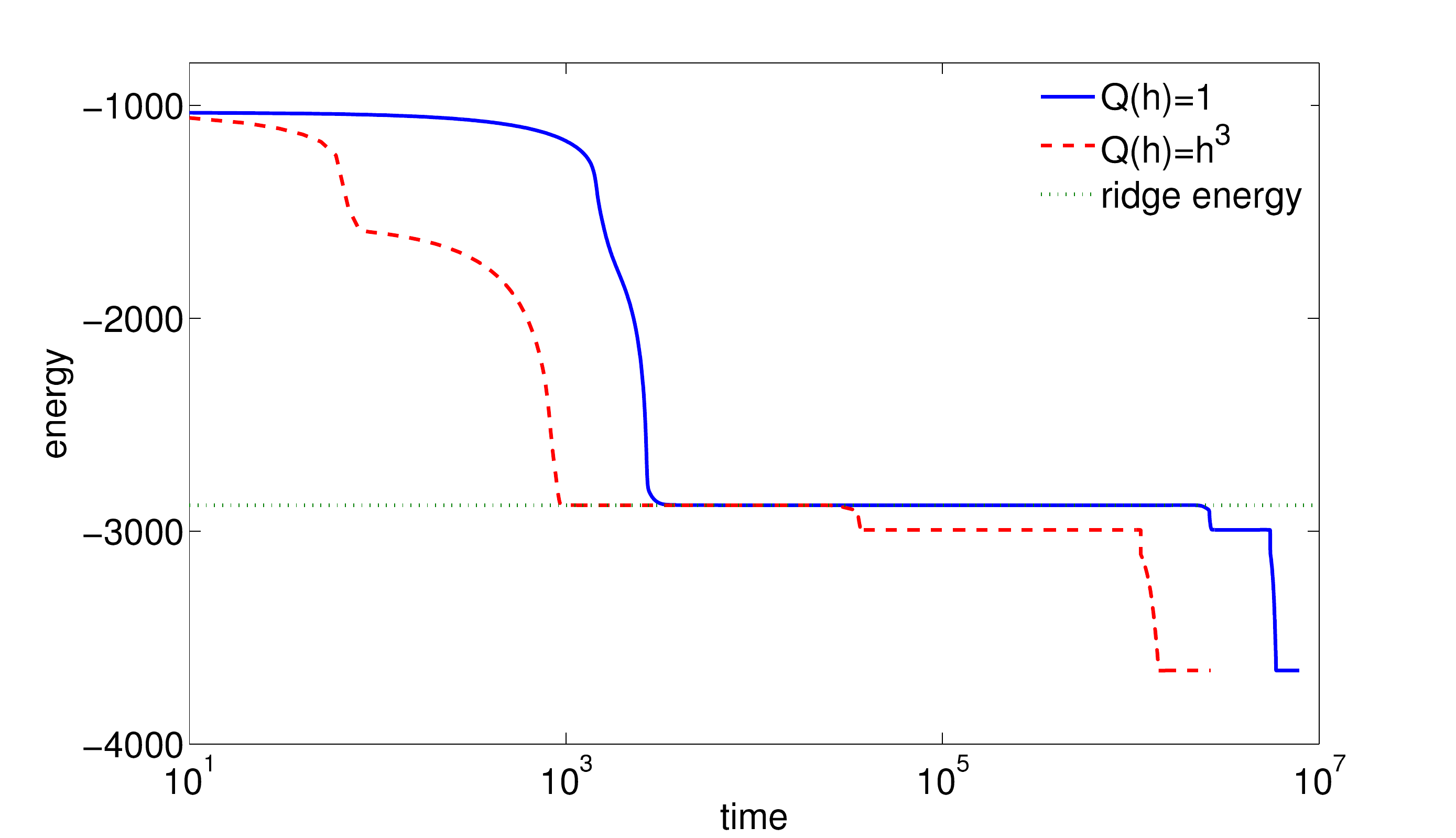}
 \end{tabular}
 \caption{Snapshots from a direct time simulation of the thin film
   equation~(\ref{eq:tfe}) with $\bar{h} = 2.8$, $l_s=0.03$,
   $\rho=0.5$, $x_A=0.3$, and $L_{per}=50$. Panels a), b), c) show
   snapshots at times $t=10^4$, $t=5\cdot 10^4$ and $t=1.5\cdot 10^6$,
   respectively. The borders of the more wettable stripes (MWS) is
   indicated by dashed (black) lines. Panel d) shows time series of
   the energy $F(t)$ for transport by surface diffusion (mobility $Q(h)\sim 1$,
   solid blue line) and by convection ($Q(h)\sim h^3$, dashed red line).} 
 \label{fig:w50}
\end{figure}

However, for larger initial film heights $\bar{h}$ the final
  bulge configuration itself (shown in Fig.~\ref{fig:w100} d)) can be
  unstable. Fig.~\ref{fig:w50} gives in panels a)-c) three snapshots
  from the time evolution for $\bar{h} = 2.8$ with convective mobility
  $Q(h) = h^3$, while Fig.~\ref{fig:w50}~d) shows the evolution of the
  energy $F(t)$ for convective transport (red dashed line) and for
  surface diffusion (blue solid line). As in the previous cases, the
  evolution visits similar unstable steady states until a state with
  two bulges is formed. However, this state does this time not only
  evolve via a coarsening event towards a single bulge, but at the
  same time forms a liquid bridge to the next MWS (the system is
  periodic in $y$-direction).  The final liquid bridge state is shown
  in~Fig.~\ref{fig:w50}~c).  However, the occurance of this final
  configuration strongly depends on the domain size and periodicity of
  the wettability pattern. A detailed investigation of this aspect is
  not within the scope of the current study and will be pursued
  elsewhere.

\section{Conclusion}
\label{sec:conc}

We have employed a mesoscopic continuum model to analyse the
redistribution stage of deposition experiments where organic molecules
are deposited by vapour deposition on a solid substrate with
stripe-like wettability patterns. In particular, we have investigated
the stability of transversally invariant ridges located on the more
wettable parts of a stripe pattern by means of a linear stability
analysis of steady one-dimensional profiles that correspond to the
transversally invariant ridges on two-dimensional
substrates. Employing a technique based on continuation procedures has
allowed us to perform the linear analysis very efficiently.  In
particular, we have studied the influence of the wettability contrast,
the mean film thickness (ridge volume), and the geometry of the stripe
pattern.

We have found that there exist two different instability modes that
on the one hand result at large ridge volume in the formation of bulges that spill from the more wettable
stripes onto the less wettable bare substrate and on the one hand result at small ridge volume in the formation of
small droplets located on the more wettable stripes, respectively. The
different modes are reflected in qualitatively different eigenmodes
that have their maxima in the regions of the strongest wettability
gradient and at the stripe centre, respectively. They can be
identified as a Rayleigh-Plateau instability and a surface instability
as in spinodal dewetting, respectively.

In the parameter plane spanned by mean film height $\bar{h}$ (liquid
volume) and wettability contrast $\rho$ one sees that a clear
distinction between the two modes appears at some wettability contrast
$\rho_c$. For $\rho>\rho_c$ two unstable film thickness ranges exist with
a stable range of film thicknesses in between. The critical $\rho_c$ decreases with
increasing steepness of the transition in wettability. In the opposite
limit of a sinusoidal wettability pattern, i.e., when the transition
zone is as wide as the period, no such distinction of two instability
modes exists.

Further, we have investigated the role of the analysed transversally
invariant ridge states in the course of the fully nonlinear time
evolution starting from a deposited homogeneous film of molecules
towards the final bulge or drop geometry. This has allowed us to
assess whether a detailed analysis of ridge stability is meaningful in
the context of deposition experiments with striped
substrates. We have found that even linearly unstable ridge
  states form the first 'organising center' of the time evolution.
  This is because they (as well as all other unstable states)
  represent saddles in the state space of the system. One may say that
  they 'attract' time evolutions from the majority of directions in
  the infinite-dimensional space of system states and later 'expel' 
  them into one or a few directions and with rates that both can be
  determined by a linear stability analysis of the unstable states.
In particular, the evolution that starts at a
flat film with small perturbations is first attracted by the ridge
state (the corresponding time scale is related to both, the most unstable
eigenmode of the flat film and the slowest stable eigenmode of the
ridge state. After 'visiting' the transversally invariant ridge state the evolution is expelled along
the single unstable direction in state space. It corresponds to the
most unstable eigenmode of the ridge state (the corresponding time scale is
the inverse of the growth rate of this eigenmode). The importance
of the ridge state and also of the other unstable states (as, e.g., multidrop states) for the
time evolution can be clearly seen in the formation of 'energy
plateaus' in all the plots that give the dependence of the free energy
on time. Each plateau represents one of the unstable steady state. This
observation is also important for other gradient dynamics systems as it indicates how
important it is to understand the \textit{complete} solution structure of a
system in order to control its time evolution. It is further to expect
that the unstable steady states are the ones that are most easily
stabilized through imposed controls - be it spatial patterning or
feedback.

The conclusion in the previous paragraph is strengthened by our finding
that the relevance of the unstable states holds for all the different 
dominant transport mechanisms that we have investigated: the same energy
plateaus appear independently of the chosen mobility, however, we have
found that their relative duration depends on the particular chosen transport
behaviour. This agrees with the fact discussed above that the steady
states do not depend on mobilities, but that the time constants and
eigenmodes do depend on them.
As particular examples for the influence of the transport mechanism of
redistribution on the evolution pathway, we have compared a standard
long-wave hydrodynamic model in the case of no slip at the substrate
(purely convective transport, cubic mobility) with diffusive transport
via diffusion in the film bulk (linear mobility) and at the film
surface (constant mobility).  All the resulting long-wave evolution
equations are gradient dynamics models on the identical underlying
free energy (interface Hamiltonian) just with different mobility
functions. As a result, one may now reconsider the film height
evolution equation and combine thin film hydrodynamics for relatively thick layers and dynamical density
functional theory for relatively thin layers by combining the
corresponding mobilities\cite{noteDiff}. This will allow one to investigate the
cross-over between different dominant transport mechanisms depending
on various parameters.

Based on our investigations we conclude more specifically that the
different experimentally observed equilibrium morphologies do not
result from the different transport mechanisms in the liquid or solid
state as thought before. Instead, we have found that the observed flat
and bulged deposits result from different balances of interface
energies and wetting energies. The influence of bulk elastic energies
may also play a role that shall be further investigated in the future.

\appendix

\section{Appendix: Implementation in AUTO-07p} \label{App:AUTO-Eq}

For the treatment with the continuation toolbox AUTO-07p, we have to
transform Eqs.~\eqref{eq:tfe_stat} and \eqref{eq:eval} into a system
of first order ODEs on the interval $[0,1]$. To this end, we first
define the independent variable $\xi:=x/L$ with $L$ denoting the physical
domain size. Next, we introduce the variables
\begin{align}
  u_1(\xi) &= h_0(L\xi) - \bar h~, \\
  u_2(\xi) &= \l.\frac{\md h_0}{\md x}\r|_{x=L\xi}~, \\
  u_3(\xi) &= h_1(L\xi)~, \\
  u_4(\xi) &= \l.\frac{\md h_1}{\md x}\r|_{x=L\xi}~, \\
  u_5(\xi) &= \l.\frac{\md^2 h_1}{\md x^2}\r|_{x=L\xi}~, \\
  u_6(\xi) &= \l.\frac{\md^3 h_1}{\md x^3}\r|_{x=L\xi}~, \\
  u_7(\xi) &= L\xi~.
\end{align}
Here, $\bar h$ denotes the mean film thickness.
With the notation $\dot u_i(\xi) = \md u_i(\xi)/\md \xi$, the system of first order ODEs reads
\begin{align}
  \dot u_1 = &Lu_2 \\
  \dot u_2 = &-L[\varPi(\bar h + u_1,u_7) + C] \\
  \dot u_3 = &Lu_4 \\
  \dot u_4 = &Lu_5 \\
  \dot u_5 = &Lu_5 \\
  \dot u_6 = &L\,\Big\{-\frac{\beta u_3}{Q_0}  + q^2u_5 -\partial_x^2(h_1\partial_h \varPi_0) \nonumber \\
&-\frac{\partial_x Q_0}{Q_0}\,\left[ (u_6-q^2u_4 +\partial_x ( h_1\partial_h \varPi_0)\right] \nonumber\\
&+q^2\,\left[ u_5-q^2u_3 + \varPi'(\bar{h} + u_1) u_3\right]  \Big\} \\
  \dot u_7 = L
\end{align}
with $Q_0 = Q(u_1 + \bar h)$ and
\begin{align}
  \partial_x (h_1 \partial_h \varPi_0) = &~\varPi''(\bar h + u_1,u_7)u_2u_3 + \varPi'(\bar h + u_1,u_7)u_4 \nonumber \\
  &+ \varPi_x'(\bar h + u_1,u_7)u_3 ~,\\
  \partial_x^2 (h_1 \partial_h \varPi_0) = &~\varPi'''(\bar h + u_1,u_7) u_2^2u_3 \nonumber \\
  &+ \varPi''(\bar h + u_1,u_7)(\partial_x^2 h_0)u_3 \nonumber \\
  & + 2 \varPi''(\bar h + u_1,u_7)u_2u_4 \nonumber \\
  & + \varPi'(\bar h + u_1,u_7)u_5 \nonumber \\
  & + 2\varPi'_x(\bar h + u_1,u_7)u_4 \nonumber \\
  & + 2\varPi''_x(\bar h + u_1,u_7)u_2u_3  \nonumber \\
  &+ \varPi'_{xx}(\bar h + u_1,u_7)u_3~.
\end{align}

% \ttuwe{there are still some symbols that are not explained. Also teh
%   use of the $'$ in expressions like $\varPi_x'$ should be mentioned.}
In the last two equations, primes denote derivatives w.\,r.\,t. $h$ while the index 
$x$ means a derivative w.\,r.\,t. $x$ at constant $h$. The equation for $x=u_7$ is 
necessary because AUTO-07p only allows for autonomous ODEs.

\begin{acknowledgements}
 This work was partly supported by the Deutsche Forschungsgemeinschaft in the framework of the
 Sino-German Collaborative Research Centre TRR 61.
 \end{acknowledgements}

% 
% 
%\bibliographystyle{achemso}
%\bibliographystyle{alpha}
%\bibliography{TF_lit_uwe}
\newcommand{\etalchar}[1]{$^{#1}$}

\end{document}